\documentclass[12pt]{article}
\pdfoutput=1

\usepackage{authblk} 
\usepackage{geometry} 
\newgeometry{vmargin={25mm}, hmargin={28mm,28mm}}

\usepackage{amsthm, amsfonts}
\usepackage{mathrsfs, mathtools}
\usepackage{physics}

\usepackage{subcaption} 
\usepackage{graphicx}

\usepackage{tikz}
\usepackage{tikz-cd}
\usepackage{pgfplots}
\pgfplotsset{compat=1.14}

\usepackage[hidelinks]{hyperref}
\hypersetup{
  colorlinks   = true, 
  linkcolor={red!50!black},
  citecolor={blue!50!black},
  urlcolor={blue!80!black}
  }
  \usepackage{cleveref}
  \usepackage{natbib}
  
\PassOptionsToPackage{dvipsnames}{xcolor}
\usepackage[dvipsnames]{xcolor}
\usepackage{multirow} 
\usepackage{relsize} 
\usepackage{siunitx} 
\usepackage{float}
\usepackage{ulem}

\title{\LARGE{\textbf{A deep learning model for inter-fraction head and neck anatomical changes}}}

\author[1,3]{Tiberiu Burlacu\thanks{Corresponding author. Contact: \texttt{t.burlacu@tudelft.nl}}}
\author[2,3,1]{Mischa Hoogeman}
\author[1,3]{Danny Lathouwers}
\author[1,3]{Zolt\'{a}n Perk\'{o}}

\affil[1]{\footnotesize{Delft University of Technology, Faculty of Applied Sciences, Delft, The Netherlands}}
\affil[2]{\footnotesize{Erasmus MC Cancer Institute, University Medical Center Rotterdam, Department of Radiotherapy, Rotterdam, The Netherlands}}
\affil[3]{\footnotesize{HollandPTC consortium\footnote{HollandPTC consortium – Erasmus Medical Center, Rotterdam, Holland Proton Therapy Centre, Delft, Leiden University Medical Center (LUMC), Leiden and Delft University of Technology, Delft, The Netherlands}, Delft, The Netherlands}}

\date{\small{\today}}

\begin{document}

\maketitle

\begin{abstract}

  \textbf{Objective:} To assess the performance of a probabilistic deep learning based algorithm for predicting inter-fraction anatomical changes in head and neck patients.

  \textbf{Approach:} A probabilistic daily anatomy model for head and neck patients (DAM\textsubscript{HN}) is built on the variational autoencoder architecture. The model approximates the generative joint conditional probability distribution of the repeat computed tomography (rCT) images and their corresponding masks on the planning CT images (pCT) and their masks. The model outputs deformation vector fields, which are used to produce possible rCTs and associated masks. The dataset is composed of \num{93} patients (i.e., \num{367} pCT - rCT pairs), \num{9} (i.e., \num{37} pairs) of which were set aside for final testing. The performance of the model is assessed based on the reconstruction accuracy and the generative performance for the set aside patients. 

  \textbf{Main results:} The model achieves a DICE score of \num{0.92} and an image similarity score of \num{0.65} on the test set. The generated parotid glands volume change distributions and center of mass shift distributions were also assessed. For both, the medians of the distributions are close to the true ones, and the distributions are broad enough to encompass the real observed changes. Moreover, the generated images display anatomical changes in line with the literature reported ones, such as the medial shifts of the parotids glands.  

  \textbf{Significance:} DAM\textsubscript{HN} is capable of generating realistic anatomies observed during the course of the treatment and has applications in anatomical robust optimization, treatment planning based on plan library approaches and robustness evaluation against inter-fractional changes. 

\end{abstract}

\vspace{0.25pc}
\noindent{\it Keywords}: proton therapy, deep learning, variational autoencoder, anatomy changes.
\maketitle

\section{Synthetic CT uses in Proton Therapy}

Proton Therapy (PT) has desirable dose characteristics, such as similar target coverage and lower organs at risk (OAR) doses, when compared to traditional photon based radiotherapy (RT) \citep{chenProtonPhotonRadiation2023}. However, the increased dose conformality implies an increased susceptibility to dose degradation by uncertainties such as setup errors, range errors and anatomical changes over the course of the typically month long treatment duration \citep{vankranenSetupUncertaintiesAnatomical2009}. To diminish the dose degradation, robust optimization and evaluation \citep{unkelbachRobustProtonTreatment2018} with isotropic setup and range settings \citep{liuEffectivenessRobustOptimization2013} and offline adaptive replanning \citep{deiterEvaluationReplanningIntensitymodulated2020} is performed in clinical practice. This results in a high dose region that surrounds the target, which in the case of the H\&N region where OARs are in close proximity to the target, could result in high chances of side effects. Moreover, there are certain anatomical changes that are not effectively accounted for by robust optimization taking setup and range errors into account. One proposed option \citep{waterAnatomicalRobustOptimization2018} is the inclusion of additional (synthetic) computed tomography (CT) images in the (anatomical) robust optimization process. While this provided increased target coverage and lower OAR doses for the specific H\&N patients in the cohort, compared to conventional robust optimization, it still created a high dose region surrounding the target. 

To reduce this region to its minimum and counter long and short-term inter-fraction occurring anatomical variations, online adaptive proton therapy (OAPT) has been proposed. In this workflow, a new CT is acquired for each fraction and within a short time a new fully re-optimized plan is generated. The resulting plan would only need minimal robustness settings to counter the effects of range uncertainties, machine related setup uncertainties and remaining intra-fraction uncertainties. The short time available and the limited computational resources imply that fully robust reoptimization in the online setting is still a topic of research \citep{oudFastRobustConstraintbased2024}. The plan library (PL) approach was proposed as an intermediate solution \citep{oudOnlineAdaptivePlan2022,vandeschootDosimetricAdvantagesProton2016}. This approach used the planning CT image to generate multiple plans with varying robustness settings. On the given day, it administers an appropriately chosen plan, therefore resulting in NTCP reductions or sometimes in increased robustness that ensures adequate target coverage. In this approach, synthethic CT images can be used to expand the pre-compiled library of plans, by generating optimal plans for the future patient anatomies predicted by the model. Assuming that in the limited time in which the patient is on the treatment table, the original treatment plan is adapted or refined, synethic CT images can prove useful for plan QA. Specifically, several CT images with associated truly optimal plans, could be generated a priori. On the given day, a fast dosimetric check can be performed between the adapted and refined plan and the truly optimal pre-generated plan. 

Thus, models of inter-fractional anatomical changes have applications in several PT related workflows such as anatomical robust optimization, plan quality assurance in OAPT or expanding the plan library approach. Multiple approaches to synthethic CT generation have been employed, such as principal component analysis (PCA) or deep learning. An overview of the different possible approaches is given by the work of \citep{smoldersDiffuseRTPredictingLikely2024}. Deep learning models have been shown to outperform PCA based ones in the case of prostate anatomies \citep{pastor-serranoProbabilisticDeepLearning2023} and denoising diffusion probabilistic models \citep{smoldersDiffuseRTPredictingLikely2024} were successfully applied for artificial CT generation for the H\&N site where they were additionally shown to increase robustness to anatomical changes. This work builds upon the previous publication of \citep{pastor-serranoProbabilisticDeepLearning2023} on a generative deep learning daily anatomy model (DAM) for prostate inter-fractional anatomical changes. The model architecture and the data processing pipeline are changed and thereafter applied to a H\&N radiotherapy cohort. The model is referred to from here on as DAM\textsubscript{HN}. Section \ref{sec:model_architecture} details the probabilistic framework of the model. Section \ref{sec:dataset_info} provides details on the dataset generation and the specific architecture configuration used for training. Section \ref{sec:results} contains the results and their discussion. The performance of the model was assessed via several tests. The results of a reconstruction accuracy test are shown in Subsection \ref{subsec:reconstruction_accuracy}. The generative performance was assessed in terms of the model's capability to predict realistic anatomical changes. To this end, an overview of the typical changes in head and neck patients reported by literature studies is given in Subsection \ref{subsec:literature_anatomical_changes}. The anatomical changes present on the training set are discussed in Subsection \ref{subsec:train_set_changes}. Subsection \ref{subsec:gen_performance} presents and discusses the anatomical changes predicted by the model. Subsection \ref{subsec:diffuse_rt} compares these anatomical changes with the ones presented in the recently published denoising diffusion probabilistic models DiffuseRT model \citep{smoldersDiffuseRTPredictingLikely2024}. Lastly, a latent space analysis is presented in Subsection \ref{subsec:latent_space_analysis}. Section \ref{sec:conclusion} concludes this work and discusses some improvement points.
\section{Model architecture}
\label{sec:model_architecture}

This section provides only the main details of this model's architecture. An in-depth exposition can be found in \citep{pastor-serranoProbabilisticDeepLearning2023}. The patient anatomy at a certain point in time is described by the CT image and the associated RT structures (masks), which are taken as random variables. On the planning CT image (pCT), an image with $N$ voxels is denoted by $\vb{x} \in \mathbb{R}^{N}$ and the corresponding structures (pM) are denoted by $\vb{s}_x \in \mathbb{R}^N$. On the repeat CT images (rCTs), the image is denoted by $\vb{y} \in \mathbb{R}^N$ and the corresponding masks (rM) by $\vb{s}_y \in \mathbb{R}^N$. 

The presence of anatomical deformations over the course of treatment, e.g., the systematic medial translation of the lateral regions of the parotid glands, the shrinkage of the parotid and submandibular glands \citep{fiorentinoParotidGlandVolumetric2012}, the change in the parotid shape from convex to flat or concave \citep{santosMorphologyVolumeDensity2020a} and the center of mass (COM) shifts towards the medial side \citep{vasquezosorioLocalAnatomicChanges2008} motivates the existence of an unknown generative joint conditional probability distribution $P^*(\vb{y}, \vb{s}_y | \vb{x}, \vb{s}_x)$ of the voxel CT HU values $\vb{y}$ and the structure masks $\vb{s}_y$ conditioned on the planning CT $\vb{x}$ and structures $\vb{s}_x$. If such a distribution were known, given a new pCT and pM, it could be sampled to generate future possible anatomies, denoted by $\vb{y}$ and $\vb{s}_y$. In general it is impossible to find such a distribution, and a good approximation $P_{\vb*{\theta}}(\vb{y}, \vb{s}_y | \vb{x}, \vb{s}_x)$ is sought instead. The distribution $P_{\vb*{\theta}}(\vb{y}, \vb{s}_y | \vb{x}, \vb{s}_x)$ is parametrized by a vector of parameters $\vb*{\theta}$ that is learned during training. 

The dataset $\mathbb{D}$ consists of elements $\vb{s}^i \in \mathbb{R}^{4N}$, which are the concatenation of a given pCT and rCT and their associated structures, i.e., $\mathbb{D} = \{\vb{s}^i=(\vb{x}^i, \vb{s}_x^i, \vb{y}^i, \vb{s}_y^i) \,| \, i = 1, \ldots, N_D \}$ with $N_D$ the number of elements in the dataset. Moreover, the dataset $\mathbb{D}$ is assumed to be independent and identically distributed (i.i.d.). As the dataset $\mathbb{D}$ is i.i.d., the log-probability assigned to the data is 
\begin{equation}
    \label{eq:log_P_dataset}
    \log P_{\vb*{\theta}}(\mathbb{D}) = \displaystyle\sum_{\vb{s} \in \mathbb{D}} \log P_{\vb*{\theta}}(\vb{s}).
\end{equation}
The framework of Maximum Likelihood (ML) searches for the parameters $\vb*{\theta}$ that maximize the sum, or equivalently the average, of the log-probabilities assigned to the data by the model in Equation \ref{eq:log_P_dataset} \citep{kingmaIntroductionVariationalAutoencoders2019}. 

\begin{figure}[H]
    \centering
    \includegraphics[width=1.0\linewidth]{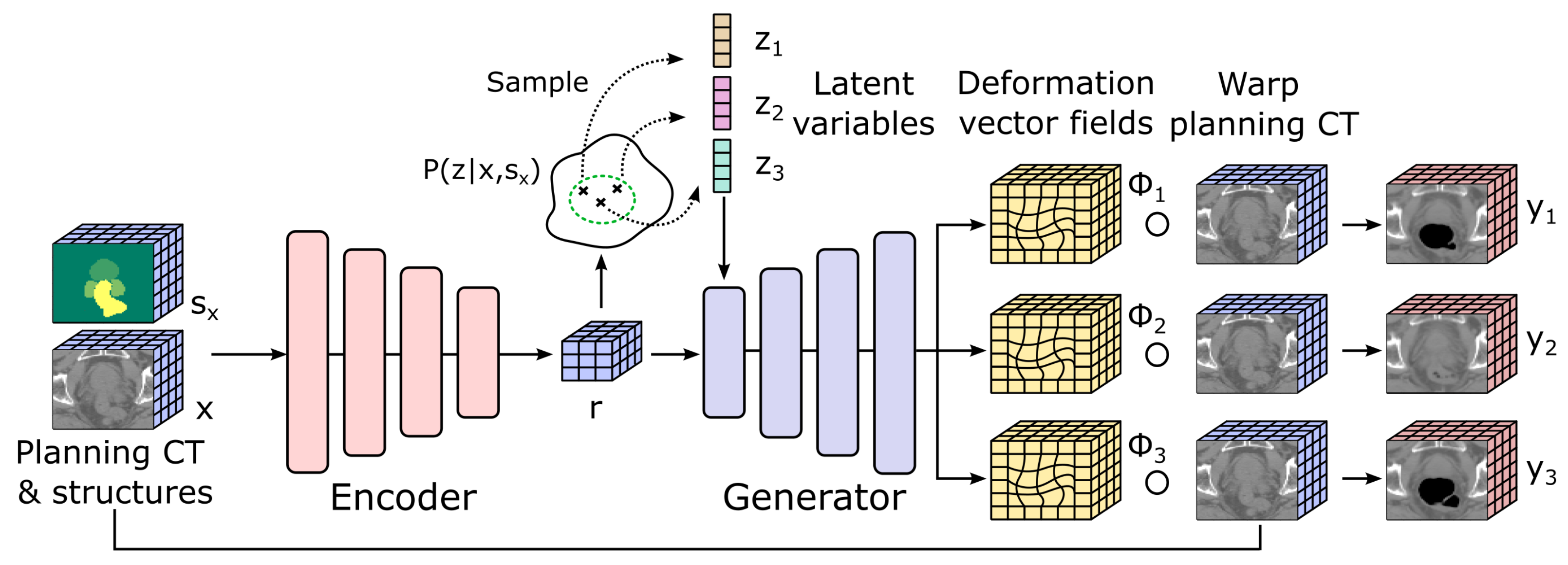}
    \caption{The proposed variational autoencoder model. The Encoder represents the down-sampling part of a U-net that computes the parameters of $P(\vb{z} | \vb{x}, \vb{s}_x)$. The up-sampling part of the U-net, denoted by Generator, takes samples from $P(\vb{z} | \vb{x}, \vb{s}_x)$ together with a reduced representation of the inputs $\vb{x}, \vb{s}_x$ and computes artificial CT images $\vb{y}, \vb{s}_y$. Figure reproduced with permission from \citep{pastor-serranoProbabilisticDeepLearning2023}.}
    \label{fig:model_framework}
\end{figure}

As most explicitly parametrized generative distributions are too simplistic to model inter-fractional anatomical variations, implicitly parametrized distributions are considered instead. Therefore, a joint conditional probability distribution $P_{\vb*{\theta}}(\vb{y}, \vb{s}_y, \vb{z} | \vb{x}, \vb{s}_x)$ that also depends on latent variables $\vb{z}$ is constructed. Latent variables are variables that are not observed, and therefore they are not part of the dataset of images and associated structures. They are meant to encode (represent in a lower dimensional space) the information between the pCT and the rCT. The marginal distribution $P_{\vb*{\theta}}(\vb{y}, \vb{s}_y | \vb{x}, \vb{s}_x)$ over the observed variables $\vb{y}, \vb{s}_y$ is recovered by marginalizing, namely
\begin{align}
    P_{\vb*{\theta}}(\vb{y}, \vb{s}_y | \vb{x}, \vb{s}_x)  
    &= \int \dd \vb{z} \,
    P_{\vb*{\theta}}(\vb{y}, \vb{s}_y, \vb{z} | \vb{x}, \vb{s}_x) \label{eq:marginal_p_data}\\
    &= \int \dd \vb{z} \,
    P_{\vb*{\theta}}(\vb{y}, \vb{s}_y | \vb{z}, \vb{x}, \vb{s}_x) 
    P_{\vb*{\theta}}(\vb{z} | \vb{x}, \vb{s}_x) \nonumber. 
\end{align}
This is also referred to as the (single datapoint) marginal likelihood, or model evidence, when taken as a function of $\vb*{\theta}$ \citep{ghojoghFactorAnalysisProbabilistic2022}. The distribution $P_{\vb*{\theta}}(\vb{z} | \vb{x}, \vb{s}_x)$ is called the prior distribution, which in the case of this work is taken as a multivariate Normal distribution with mean and variance that depend on the pCT and pM and on the vector of learned parameters $\vb*{\theta}$, namely
\begin{align}
    P_{\vb*{\theta}}(\vb{z} | \vb{x}, \vb{s}_x) = \mathscr{N}(\vb{z}; \vb*{\mu}_{\vb*{\theta}}(\vb{x}, \vb{s}_x), \Sigma_{\vb*{\theta}}(\vb{x}, \vb{s}_x)).
\end{align}
The dependence of the parameters of the prior distribution on the pCT and pM results in a different distribution for each patient (insofar as a patient is identified with a single image). The mean $\vb*{\mu}_{\vb*{\theta}}(\vb{x}, \vb{s}_x)$ and the covariance matrix $\Sigma_{\vb*{\theta}}(\vb{x}, \vb{s}_x)$ are computed in the down-sampling part of a U-net neural network and the parameters $\vb*{\theta}$ of the prior are the weights of the encoder, as illustrated in Figure \ref{fig:model_framework}. 

The up-sampling part of the U-net, denoted by Generator in Figure \ref{fig:model_framework}, outputs a deformation vector field (DVF) $\Phi : \mathbb{R}^{N \cross 3} \rightarrow \mathbb{R}^{N \cross 3}$ used to map coordinates $\vb{p} \in \mathbb{R}^3$ between images. The DVF $\Phi$ is used to obtain the prediction of the model $\vb{y} = \Phi \circ \vb{x}$ \citep{jaderbergSpatialTransformerNetworks2016}. Based on work by \citep{krebsLearningProbabilisticModel2019}, the distribution $P_{\vb*{\theta}}(\vb{y}, \vb{s}_y | \vb{z}, \vb{x}, \vb{s}_x)$ (referred to as the likelihood) is taken as a function of the normalized cross-correlation (NCC) between the ground truth image $\hat{\vb{y}}$ and the predicted image $\vb{y}$ with an additional scaling factor $w_{NCC} \in \mathbb{R}^{+}$, namely 
\begin{align}
    P_{\vb*{\theta}}(\vb{y}, \vb{s}_y | \vb{z}, \vb{x}, \vb{s}_x) 
    = \exp\left(- w_{\mathrm{NCC}} \mathrm{CC}(\vb{y}, \hat{\vb{y}}) \right),
    \label{eq:NCC}
\end{align}
where the $\mathrm{CC}$ term is defined as
\begin{align}
    \mathrm{CC}(\vb{y}, \hat{\vb{y}}) = \mathlarger\displaystyle\sum_{\vb{p} \in \Omega} 
    \frac{\left[\displaystyle\sum_{i=1}^{n^3} \left(\hat{\vb{y}}(\vb{p}_i) - \hat{w}(\vb{p}) \right) \left(\vb{y}(\vb{p}_i) - w(\vb{p})\right)\right]^2}{\left[\displaystyle\sum_{i=1}^{n^3} \left(\hat{\vb{y}}(\vb{p}_i) - \hat{w}(\vb{p}) \right) \right] \left[\displaystyle\sum_{i=1}^{n^3} \left(\vb{y}(\vb{p}_i) - w(\vb{p})\right)\right]},
\end{align}
and $w(\vb{p})$ and $\hat{w}(\vb{p})$ are the local mean over a small cube with $n$ of voxels of the generated and true images, namely
\begin{align*}
    w(\vb{p}) = \frac{1}{n^3} \displaystyle\sum_{j=1}^{n^3} \vb{y}(\vb{p}_j) \text{, and }
    \hat{w}(\vb{p}) = \frac{1}{n^3} \displaystyle\sum_{j=1}^{n^3} \hat{\vb{y}}(\vb{p}_j).
\end{align*}
The parameters $\vb*{\theta}$ of the likelihood distribution $P_{\vb*{\theta}}(\vb{y}, \vb{s}_y | \vb{z}, \vb{x}, \vb{s}_x)$ are the weights of the Generator and thus, the vector $\vb*{\theta}$ holds the weights of both the Encoder and the Generator networks. 

The main difficulty of this proposed framework is that the marginal probability of the data, or the model evidence, given in Equation \ref{eq:marginal_p_data} is intractable due to not having an analytic solution or an efficient estimator. In turn, this makes optimization of such a model computationally expensive.


\subsection{Learning the optimal parameters}

To overcome the previously mentioned intractability of the framework, the posterior distribution $P_{\vb*{\theta}}(\vb{z}| \vb{y}, \vb{s}_y, \vb{x}, \vb{s}_x) $ is approximated by a multivariate Normal distribution $Q_{\vb*{\psi}}(\vb{z} | \vb{y}, \vb{s}_y, \vb{x}, \vb{s}_x)$ parametrized by a vector of parameters $\vb*{\psi}$ with mean and variance that depend on both the planning and repeat images and masks, namely
\begin{align}
    Q_{\vb*{\psi}} = \mathscr{N}(\vb{z}; \vb*{\mu}_{\vb*{\psi}}(\vb{x}, \vb{s}_x, \vb{y}, \vb{s}_y), \Sigma_{\vb*{\psi}}(\vb{x}, \vb{s}_x, \vb{y}, \vb{s}_y)). 
\end{align}

\begin{figure}[H]
    \centering
    \includegraphics[width=\linewidth]{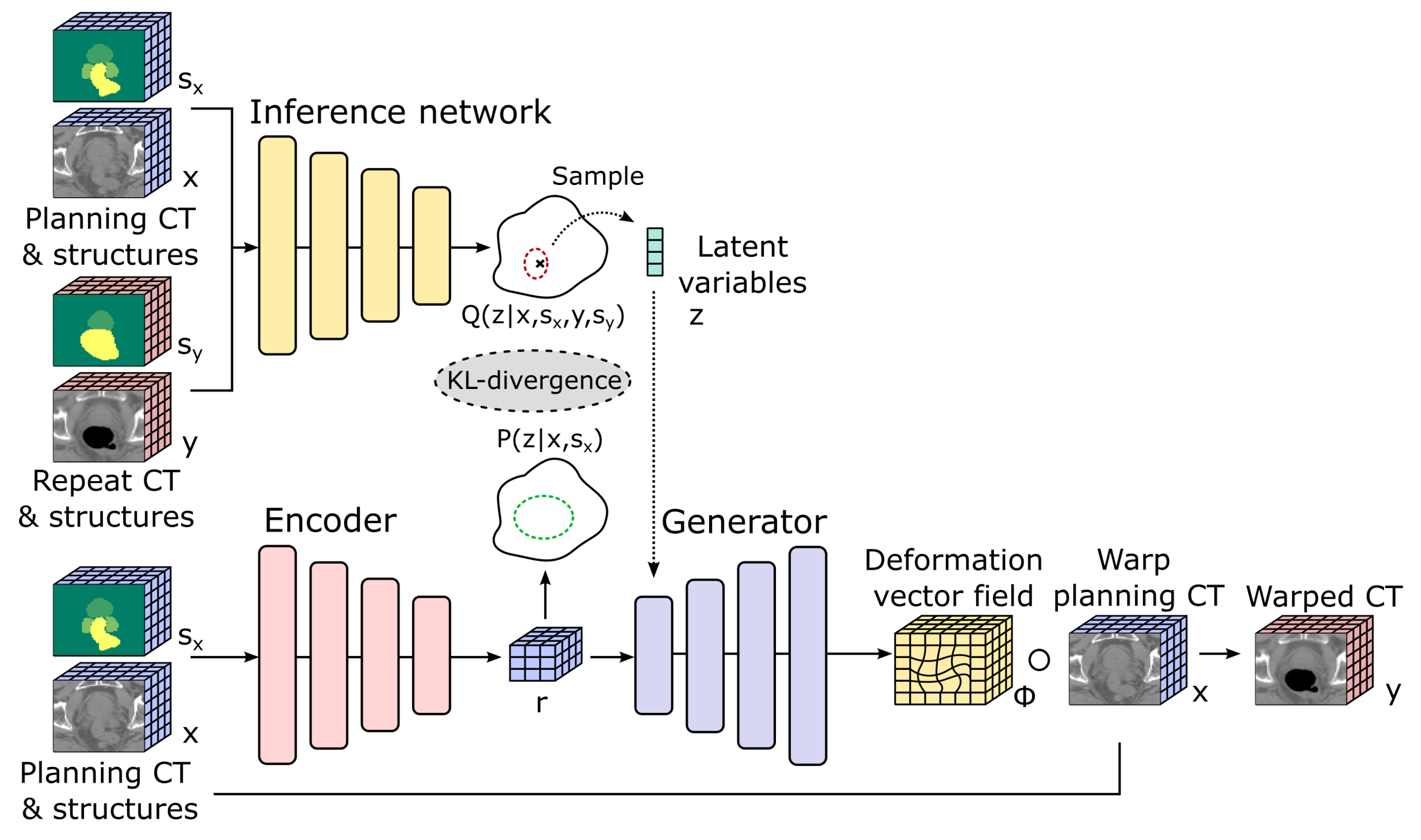}
    \caption{Architecture for finding the optimal parameters $\vb*{\theta}$, $\vb*{\psi}$ of the network. Figure reproduced with permission from \citep{pastor-serranoProbabilisticDeepLearning2023}.}
    \label{fig:model_architecture}
\end{figure}

The parameters $\vb*{\psi}$ are the weights of the down-sampling part of a U-net, referred to as Inference network at the top of Figure \ref{fig:model_architecture}.

Regardless of the choice of the approximated posterior distribution $Q_{\vb*{\psi}}$, the log-likelihood of the data can be written as
\begin{align}
    \log P_{\vb*{\theta}}(\vb{y}, \vb{s}_y | \vb{x}, \vb{s}_x) &= 
    \mathbb{E}_{\, Q_{\vb*{\psi}}} \left[ \log P_{\vb*{\theta}}(\vb{y}, \vb{s}_y | \vb{x}, \vb{s}_x) \right] \nonumber \\
    &= \mathbb{E}_{\, Q_{\vb*{\psi}}} 
    \left[ \log 
        \frac{P_{\vb*{\theta}}(\vb{y}, \vb{s}_y, \vb{z} | \vb{x}, \vb{s}_x)}{Q_{\vb*{\psi}}(\vb{z} | \vb{y}, \vb{s}_y, \vb{x}, \vb{s}_x)}
    \right] \label{eq:ELBO_deriv} \\
    &+ D_{KL}\left(Q_{\vb*{\psi}}(\vb{z} | \vb{y}, \vb{s}_y, \vb{x}, \vb{s}_x) || P_{\vb*{\theta}}(\vb{z} | \vb{y}, \vb{s}_y, \vb{x}, \vb{s}_x) \right). \label{eq:D_KL_approx_posterior}
\end{align}
The $D_{KL}$ term in Equation \ref{eq:D_KL_approx_posterior} defines the Kullback-Leibler divergence between the approximated posterior distribution and the true posterior distribution. The term is non-negative, measures the distance between the shapes of the two distributions, and is zero if, and only if, the approximated posterior equals the true posterior. The expectation term in Equation \ref{eq:ELBO_deriv}, defines the evidence lower bound (ELBO) as 
\begin{equation*}
    \mathcal{L}_{\vb*{\theta}, \vb*{\psi}} = \mathbb{E}_{\, Q_{\vb*{\psi}}}
    \left[
        \log P_{\vb*{\theta}}(\vb{y}, \vb{s}_y, \vb{z} | \vb{x}, \vb{s}_x)
        - \log Q_{\vb*{\psi}}(\vb{z} | \vb{y}, \vb{s}_y, \vb{x}, \vb{s}_x)
    \right], 
\end{equation*} 
which can also be re-written as
\begin{equation}
    \label{eq:ELBO_def}
    \mathcal{L}_{\vb*{\theta}, \vb*{\psi}} = \mathbb{E}_{\, Q_{\vb*{\psi}}}
    \left[
        \log P_{\vb*{\theta}}(\vb{y}, \vb{s}_y | \vb{z}, \vb{x}, \vb{s}_x)
    \right]
    - D_{KL}(Q_{\vb*{\psi}}(\vb{z} | \vb{y}, \vb{s}_y, \vb{x}, \vb{s}_x) || P_{\vb*{\theta}}(\vb{z} | \vb{x}, \vb{s}_x)). 
\end{equation} 
As the $D_{KL}$ term is non-negative, it is clear that the ELBO is a lower bound on the log-likelihood of the data, i.e., 
\begin{align*}
    \mathcal{L}_{\vb*{\theta}, \vb*{\psi}} &= 
    \log P_{\vb*{\theta}}(\vb{y}, \vb{s}_y | \vb{x}, \vb{s}_x)
    - D_{KL}\left(Q_{\vb*{\psi}}(\vb{z} | \vb{y}, \vb{s}_y, \vb{x}, \vb{s}_x) || P_{\vb*{\theta}}(\vb{z} | \vb{y}, \vb{s}_y, \vb{x}, \vb{s}_x) \right) \\
    &\leq \log P_{\vb*{\theta}}(\vb{y}, \vb{s}_y | \vb{x}, \vb{s}_x).
\end{align*}
Thus, by maximizing the ELBO $\mathcal{L}_{\vb*{\theta}, \vb*{\psi}}$ from Equation \ref{eq:ELBO_def} with respect to the parameters of the model $\vb*{\theta}, \vb*{\psi}$ the marginal likelihood $P_{\vb*{\theta}}$ is approximately maximalized resulting in a better generative model and the KL divergence between the approximated posterior and the true posterior is lowered. 

To improve model performance, the ELBO is expanded with two additional terms which are included via multiplication to the likelihood from Equation \ref{eq:ELBO_def} \citep{pastor-serranoProbabilisticDeepLearning2023}. The first is a spatial regularization term, 
\begin{align}
    R(\Phi) = - \, w_{\mathrm{REG}} \displaystyle\sum_{\vb{p} \in \Omega} \norm{\grad \Phi(\vb{p})}_2,
\end{align}
where $w_{\mathrm{REG}}$ is a multiplication constant. This term penalizes large and unrealistic deformations and encourages neighboring voxels to deform somewhat similarly. This term could be a limitation in anatomical regions that involve tissue slipping (e.g., in the case of the abdomen where bowel slippage occurs while neighboring organs remain in the same place). 

The second is a segmentation regularization term using the DICE score is added, which is also multiplied by a constant $w_{DICE}$. This aims to improve the overlap between the propagated and ground truth structures, and is written as
\begin{align}
    \mathrm{DICE}(\vb{s}_y^k, \hat{\vb{s}}_y^k) 
    = 2 \, w_{\mathrm{DICE}} \, \frac{\abs{\vb{s}_y^k \cap\hat{\vb{s}}_y^k}}{\abs{\vb{s}_y^k} + \abs{\hat{\vb{s}}_y^k}}, 
    \label{eq:DICE}
\end{align}
where $k$ denotes the index of the structure present in the CT image, $k \in [1, K]$, with $K$ the total number of structures present, and $\vb{s}_y^k$ and 
$\hat{\vb{s}}_y^k$ are the k-th generated and ground truth structures respectively.

\section{Dataset generation and training details}
\label{sec:dataset_info}

The dataset was acquired from the Holland Proton Therapy Center and came from 93 H\&N patients with planning, repeat CT images and associated RT structures for each image. This resulted in 367 pCT - rCT pairs from which \SI{10}{\percent}, corresponding to 9 patients, were set aside for final testing. The remaining part was divided into \SI{5}{\percent} for validation and \SI{95}{\percent} for training. All the rCTs were rigidly registered to the pCTs using the Simple ITK library \citep{beareImageSegmentationRegistration2018} with the resulting deformation vector fields used to register the RT masks. After this, all scans were interpolated to a \numproduct{2x2x2} \unit{\mm} grid and cropped around the center of mass of the present RT masks (the left and right parotid glands) into a shape of \numproduct{96x96x64} voxels. This resulted in volumes of \numproduct{192x192x128} \unit{\mm^3} which were found to adequately cover the anatomical regions of interest. 

The down-sampling path of the U-net (Generator) and the Inference network were identical, and consisted of 4 blocks, where each block is composed of a 3D convolution layer, a Group Normalization layer, a rectified linear (ReLu) activation and a max pooling down-sampling operation. All convolution layers had a kernel of dimensions \numproduct{3 x 3 x 3}. The convolution layer in the first block had 16 channels while the remaining blocks had 32. At the lowest level, a last convolution with 4 channels results in the encoded volume $\vb{r} \in \mathbb{R}^{4 \cross 4 \cross 4 \cross 3}$. This volume is mapped to the means and variances via two different fully-connected layers. The up-sampling part of the U-net (Generator) concatenates the sampled latent variables to the volume $\vb{r}$ after a linear layer. Next, 7 blocks (with up-sampling as opposed to down-sampling max pooling operations) are applied, where for the first 5 the convolutional layer has 32 channels and for the last 2, the convolutional layer has 16 channels. This is followed by a last convolution with 3 channels. The model was trained using a batch size of 32, on a A40 NVIDIA GPU, for 1500 epochs with an early stopping patience of 300 epochs and the Adam with a learning rate of \num{1.0e-4}.

The constants $w_{\mathrm{NCC}}, w_{\mathrm{DICE}}, w_{\mathrm{REG}}$ together with the constant $w_{\mathrm{KL}}$ that multiplied the $D_{\mathrm{KL}}$ loss term were considered as hyparparameters to be optimized. These hyperparameters were optimized on the validation set, by choosing as validation loss the NCC from equation \ref{eq:NCC} and DICE from Equation \ref{eq:DICE} with unity weights. Thus, for a given latent space dimension, the hyperparameters were varied and the model with the lowest validation loss was chosen. This resulted in the model with $w_{\mathrm{NCC}} = 5000, \, w_{\mathrm{DICE}} = 3000, w_{\mathrm{REG}} = \num{1.0e-4}$ and $w_{\mathrm{KL}} = 1$.
\section{Results and discussion}
\label{sec:results}

This section presents and discusses the performance of the model in a series of tests. The section starts by presenting and discussing in Subsection \ref{subsec:reconstruction_accuracy} the performance of the model on the test set (a reconstruction accuracy test). Next, a baseline is set through a literature study for the expected anatomical changes in H\&N patients in Subsection \ref{subsec:literature_anatomical_changes}. The anatomical changes displayed by the training set are compared to the expectations set out by literature, in Subsection \ref{subsec:train_set_changes}. Given this framework, the generative performance of the model is presented and discussed in Subsection \ref{subsec:gen_performance}. To gain insight into the model, a latent space analysis is presented and discussed in Subsection \ref{subsec:latent_space_analysis}. Lastly, a comparison to the recent diffusion model proposed by \citep{smoldersDiffuseRTPredictingLikely2024} is given in Subsection \ref{subsec:diffuse_rt}. 

\subsection{Test set accuracy}
\label{subsec:reconstruction_accuracy}

\begin{figure}[H]
    \centering
    \includegraphics[width=\linewidth]{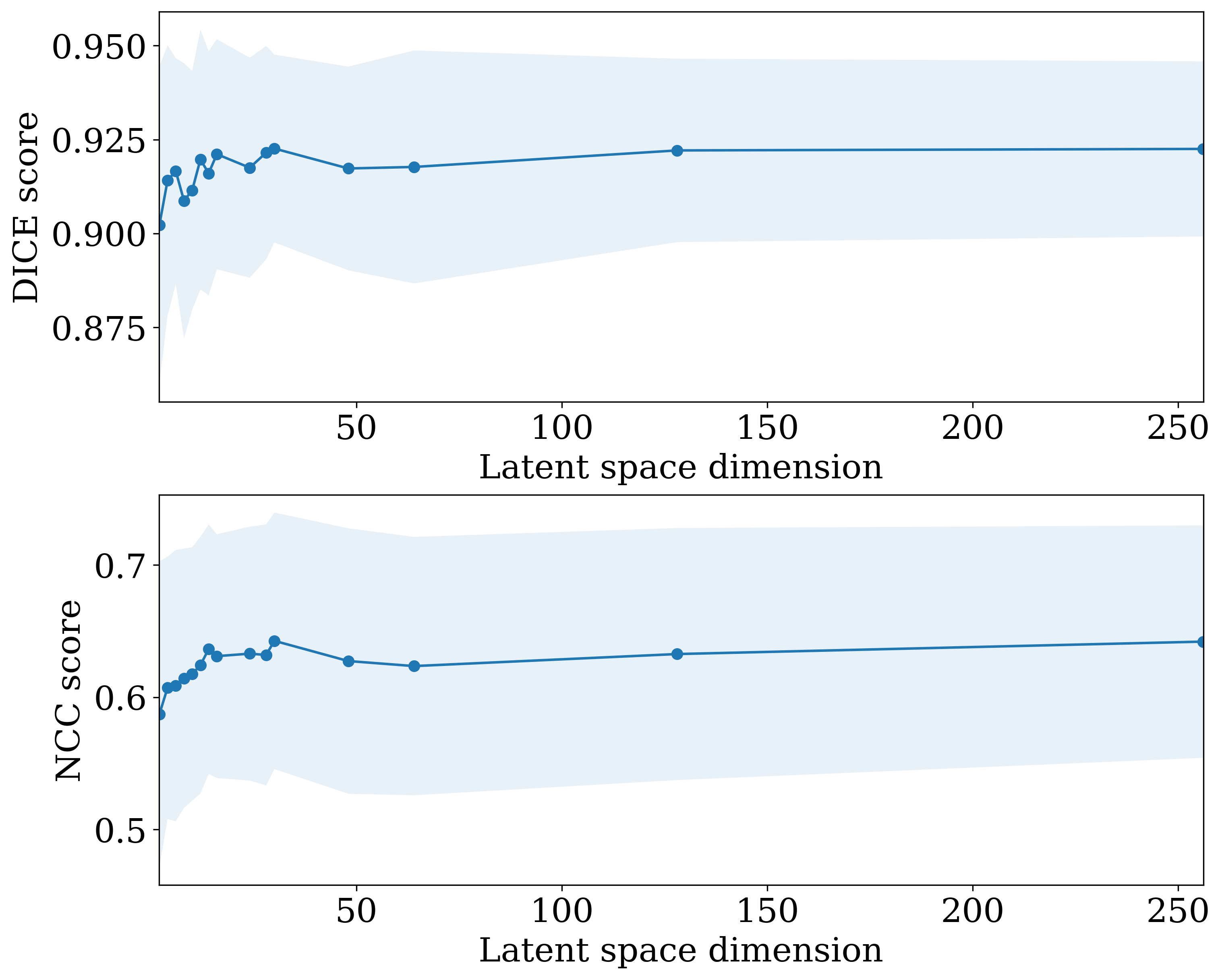}
    \caption{Reconstruction accuracy versus latent space dimensionality. The figure displays on the first row the DICE score and on the second row the NCC evaluated on the test set for models with the same hyperparameters and varying latent space dimensions.}
    \label{fig:reconstruction_accuracy}
\end{figure}

The reconstruction accuracy of the model on the test set was assessed. This test was defined by two metrics, namely the normalized cross correlation (NCC) loss from Equation \ref{eq:NCC} and the DICE loss from Equation \ref{eq:DICE}. Thus, each record in the test set (i.e., pair of pCT and rCT with associated masks) was used to generate through the inference network latent variables, which ultimately result in generated CTs and associated structures. The results were averaged over all records in the test set and the dimension of the latent space was varied between 2 and 256. The results of the two scores can be seen in Figure \ref{fig:reconstruction_accuracy}, which shows the mean of the individual scores and a band of one standard deviation around the mean. Both figures show a significant improvement in both metrics as the latent space is increased from 2 to 32, and thereafter a plateau occurring between 64 and 256. The model performs particularly well with regard to the DICE score, where it achieves a score of \num{0.9} with just 2 latent variables. The reconstruction accuracy values obtained are not directly comparable with the ones previously published in the work of \citep{pastor-serranoProbabilisticDeepLearning2023}, but exhibit the same behavior. The increased input size of this model (\numproduct{96x96x64} versus \numproduct{64x64x48}), the different anatomical site (H\&N versus prostate) and a different number of layers in the Inference, Encoder and Generator networks likely explain the need for additional latent variables to achieve good performance. 

\subsection{Expected anatomical changes}
\label{subsec:literature_anatomical_changes}

This subsection details the anatomical changes in head and neck patients that literature studies report on. An overview of these changes can be seen in Table \ref{tab:literature_anatomical_changes}. This overview was used to assess the changes observed in the training set, and therefore the changes that the DAM\textsubscript{HN} predicts.  

The work of \citep{bhideWeeklyVolumeDosimetric2010} used repeat CT scans at weeks 2, 3, 4, and 5 during radiotherapy and compared the parotids and the target at succesive time points, i.e., pretreatment with week 2, week 2 with week 3, and so on. The greatest absolute and percent reduction in the volume of the parotid glands was \SI{4.2}{\cm^3} or \SI{14.7}{\percent}, and occurred between week 0 and week 2. The absolute and percent reduction in the next two-week period was \SI{4}{\cm^3} or \SI{16}{\percent}. The study found a significant medial shift of the parotid glands through the course of treatment, starting at week 2, with the highest mean movement of the COM being \SI{2.3}{\mm} at week 4. No significant movements of the COM in the anteroposterior and the inferosuperior directions were found.

In the work of \citep{vasquezosorioLocalAnatomicChanges2008} the impact of \SI{46}{\gray} delivered to the tumor was assessed based on the planning and repeat CT images. They report that the parotids shrunk on average by \SI{14}{\percent} and that the shrinkage occurred by keeping the regions nearby to bony anatomy as an anchor. Moreover, the parotids exhibited a tendency to move inward (right parotid leftward and left parotid rightward) with the largest displacements being in the lateral and inferior regions. The region that moved the least was the medial region (partially adjacent to the bony structure). The study of \citep{barkerQuantificationVolumetricGeometric2004} found a median medial shift of \SI{3.1}{\mm} for the center of mass of the parotid glands. They observed asymmetric shifts in parotid gland surfaces, with average displacements of 1 $\pm$ 3 \unit{mm} and 3 $\pm$ 3 \unit{mm} for the medial and lateral regions of the irradiated glands, respectively. 

\begin{table}[H]
    \centering
    \resizebox{\textwidth}{!}{%
    \begin{tabular}{|lllll|}
    \hline
    \multicolumn{1}{|c}{Study} &
      \multicolumn{1}{c}{CT number} &
      \multicolumn{1}{c}{\begin{tabular}[c]{@{}c@{}}Volumetric \\ loss\end{tabular}} &
      \multicolumn{1}{c}{\begin{tabular}[c]{@{}c@{}}COM \\ shift\end{tabular}} &
      \multicolumn{1}{c|}{\begin{tabular}[c]{@{}c@{}}Morphological\\ alterations and notes\end{tabular}} \\ \hline
    \citep{barkerQuantificationVolumetricGeometric2004} &
      $\geq$ 2 &
      \begin{tabular}[c]{@{}l@{}}Median \SI{190}{\mm^{3}} per day\\ Range of \numrange{40}{840} \unit{\mm^{3}} per day\end{tabular} &
      \begin{tabular}[c]{@{}l@{}}Median \SI{3.1}{\mm} \\ Range \numrange{0}{9.9} \unit{\mm}\\ in medial direction\end{tabular} &
      \begin{tabular}[c]{@{}l@{}}Shrinkage correlated with\\ patient weight loss\end{tabular} \\ \hline
    \citep{vasquezosorioLocalAnatomicChanges2008} &
      2 &
      Average \SI{14}{\percent} &
      1 or 3 \unit{mm} &
      \begin{tabular}[c]{@{}l@{}}Bony anatomy kept as anchor\\ during shrinkage\end{tabular} \\ \hline
    \citep{bhideWeeklyVolumeDosimetric2010} &
      $\geq$ 2 &
      \begin{tabular}[c]{@{}l@{}}\SI{14}{\percent} or \SI{4200}{\mm^{3}} \\ between week 0 and 2\\ \SI{16}{\percent} or \SI{4000}{\mm^{3}} \\ between week 2 and 4\\ \SI{35}{\percent} over the course of \\ chemoradiotherapy \end{tabular} &
      \begin{tabular}[c]{@{}l@{}}\SI{2.3}{\mm} by week 4 \\ in the medial\\ direction\end{tabular} &
      \begin{tabular}[c]{@{}l@{}}COM shift insignificant \\ in the anteroposterior \\ and inferosuperior directions\end{tabular} \\ \hline
    \citep{santosMorphologyVolumeDensity2020a} &
      2 &
      \begin{tabular}[c]{@{}l@{}}Average \SI{20.5}{\percent} \\ or \SI{6560}{\mm^{3}} between CTs\end{tabular} &
      N.A. &
      \begin{tabular}[c]{@{}l@{}}Shape shift from convex \\ to concave\\ COM shift towards the\\ medial and cranial directions\end{tabular} \\ \hline
    \end{tabular}%
    }
    \caption{Overview of documented quantitative and qualitative anatomical changes in the parotid glands. The table displays the study, the number of CTs used, the reported volumetric change (absolute, relative or both), the absolute shifts in the COM and its direction and qualitative notes on the reported changes.}
    \label{tab:literature_anatomical_changes}
\end{table}

\subsection{Training set anatomical changes}
\label{subsec:train_set_changes}

\begin{figure}[H]
    \centering
    \begin{subfigure}{\linewidth}
        \includegraphics[width=\columnwidth]{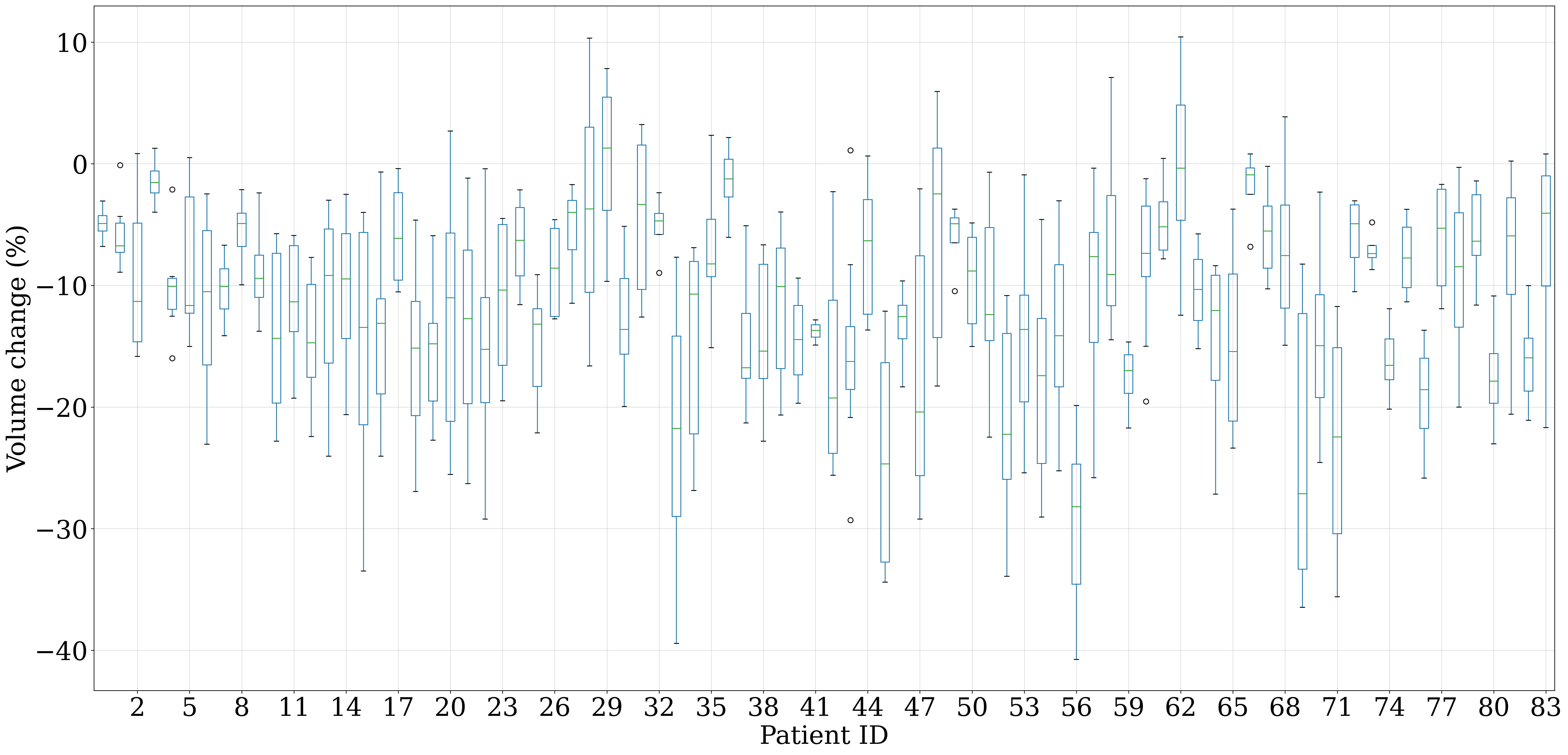}
        \caption{Patient specific box plot of relative difference in parotid volume changes.}
    \end{subfigure}
    \begin{subfigure}{\linewidth}
        \includegraphics[width=\columnwidth]{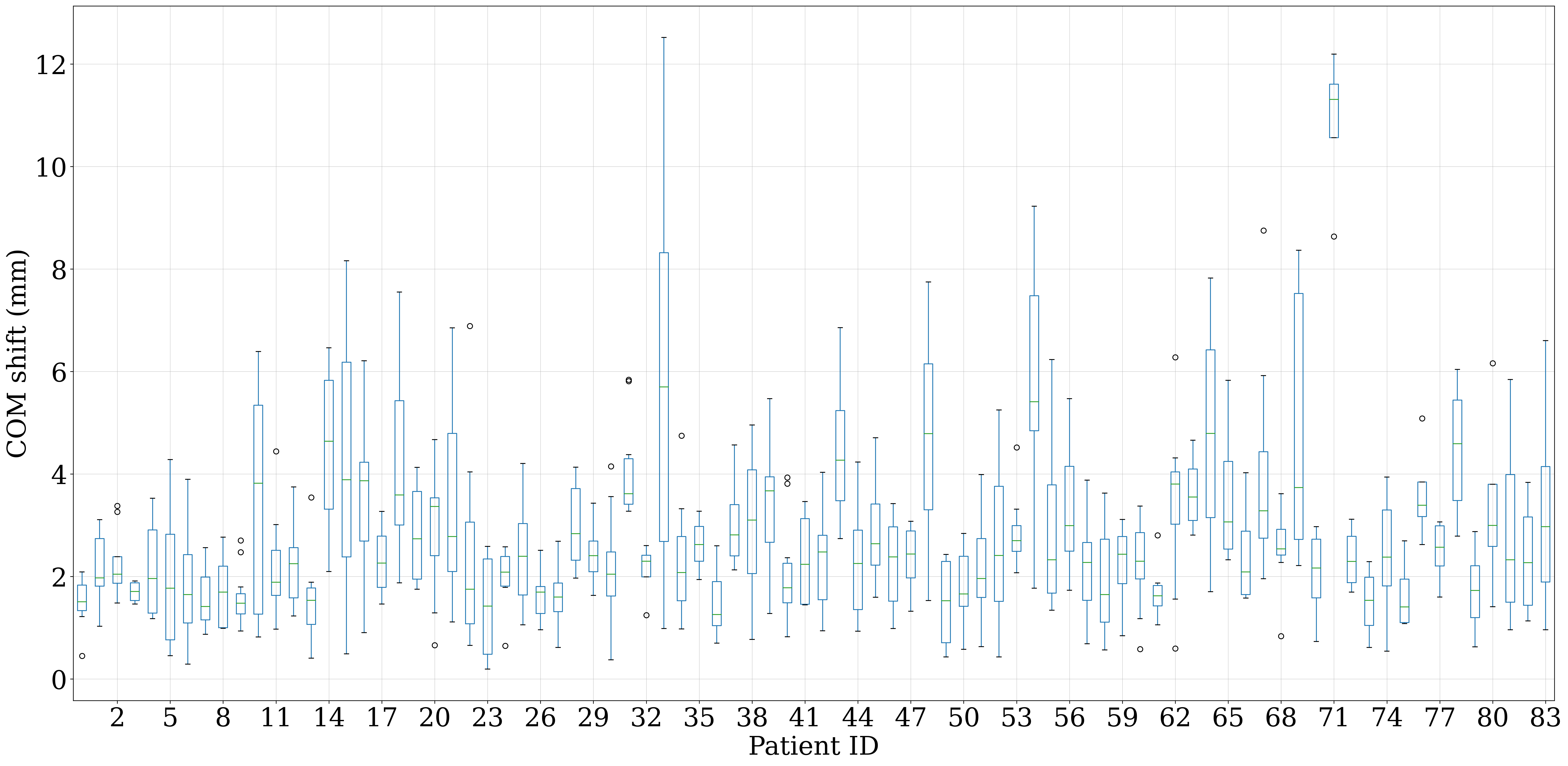}
        \caption{Patient specific box plot of COM shifts.}
    \end{subfigure}
    \caption{Training set characterization. The figures display boxplots with the $x$ axis giving the patient identifying number and the $y$ axis giving either the relative volumetric changes or the COM shifts averaged over both parotid glands.}
    \label{fig:train_set_pat_boxplot}
\end{figure}

The generative performance of the model is tied to the data provided during training in the training set. Therefore, the anatomical changes in the training set and the literature reported changes from Table \ref{tab:literature_anatomical_changes} were compared. It is worth noting that the anatomical changes presented in Subsection \ref{subsec:literature_anatomical_changes} come from studies in which uni or bilateral photon-based radiotherapy (RT) or a combination of chemotherapy and RT was delivered. In contrast, the dataset of this work comes exclusively from proton therapy patients treated with mostly bilateral fields. The training set contained anonymized data and was composed of pairs of pCTs and consecutive rCTs (pCT-rCT\textsubscript{1}, pCT-rCT\textsubscript{2}, and so on). For each such pair and patient, the volume loss and COM shift in each parotid was computed and averaged over both parotid glands. Figure \ref{fig:train_set_pat_boxplot} displays, for each patient in the training set, boxplots of the percentage parotid volume changes and parotid center of mass shifts. 

The Figure shows that the median of the volumetric loss in the parotids is $\approx$ \SI{11}{\percent} and the median of the COM shift is $\approx$ \SI{3}{\mm}. While the majority of patients have relatively unskewed volumetric change and COM shifts distributions, there are also patients (e.g., 4, 32, 49 and 66) that display skewed distributions with outliers. To facilitate comparison to previous publications, the data presented in Figure \ref{fig:train_set_pat_boxplot} is summarized in Table \ref{tab:train_set_stats} where statistics on an individual parotid level are displayed. 

\begin{table}[H]
    \centering
    \begin{tabular}{|cl|lllll|}
    \hline
                                                     &                          & \multicolumn{5}{c|}{Statistic}         \\ \hline
    \multicolumn{1}{|c|}{Organ} &
      \multicolumn{1}{c|}{Metric} &
      \multicolumn{1}{c}{Mean} &
      \multicolumn{1}{c}{SD} &
      \multicolumn{1}{c}{Min.} &
      \multicolumn{1}{c}{Median} &
      \multicolumn{1}{c|}{Max.} \\ \hline
    \multicolumn{1}{|c|}{\multirow{5}{*}{Parotid L}} & Planning volume (\unit{\mm^3})         & 35878 & 11290 & 16984  & 33280 & 83520 \\
    \multicolumn{1}{|c|}{}                           & Repeat volume (\unit{\mm^3})           & 31571 & 10161 & 12976  & 29816 & 76632 \\
    \multicolumn{1}{|c|}{}                           & Difference  (\unit{\mm^3})             & -4307 & 3880  & -30456 & -3548 & 4256  \\
    \multicolumn{1}{|c|}{}                           & Relative difference (\%) & -12   & 9     & -41    & -11   & 10    \\
    \multicolumn{1}{|c|}{}                           & COM shift (\unit{\mm})                & 3     & 2     & 0.2    & 2     & 13    \\ \hline
    \multicolumn{1}{|c|}{\multirow{5}{*}{Parotid R}} & Planning volume  (\unit{\mm^3})        & 35447 & 12568 & 11344  & 33024 & 87352 \\
    \multicolumn{1}{|c|}{}                           & Repeat volume  (\unit{\mm^3})          & 31507 & 11160 & 7496   & 29896 & 79136 \\
    \multicolumn{1}{|c|}{}                           & Difference  (\unit{\mm^3})             & -3941 & 3955  & -29112 & -3320 & 4584  \\
    \multicolumn{1}{|c|}{}                           & Relative difference (\%) & -11   & 8     & -41    & -10   & 10    \\
    \multicolumn{1}{|c|}{}                           & COM shift   (\unit{\mm})             & 3     & 2     & 0.4    & 3     & 12    \\ \hline
    \end{tabular}
    \caption{Training set statistics. The table displays for both parotid glands the mean, standard deviation, minimum, median and maximum of the volume on the planning and repeat CT images, the difference between these volumes (absolute and relative) and the center of mass shifts. }
    \label{tab:train_set_stats}
\end{table}

Specifically, the absolute volumes on the planning and repeat CT images, their difference (absolute and relative) and the COM shifts are characterized through their mean, standard deviation (SD), minimum, median and maximum. The absolute volumes of the parotids on the pCT images are a mean of \SI{35878}{\mm^3} with a range of \numrange{16984}{83520} \unit{\mm3} for the left parotid and a mean of \SI{35447}{\mm^3} with a range of \numrange{11344}{87352} \unit{\mm3} for the right parotid. Both mean parotid volumes are roughly \SI{23}{\percent} larger than the volumes reported by \citep{santosMorphologyVolumeDensity2020a}, namely \SI{28477}{\mm^3} for the left parotid and \SI{29274}{\mm^3} for the right parotid. The study of \citep{medberyVariationParotidGland2000a} also reports on the absolute volume of the parotids, where a median of \SI{25262}{\mm^3} with a range of \numrange{9225}{54080} \unit{\mm^3} is found, in close agreement with the study of \citep{santosMorphologyVolumeDensity2020a}. 

The differences between the parotid volumes in the training set are, a mean of \SI{-4307}{\mm^3} with a range of \qtyrange{-30456}{4256}{\mm^3} corresponding to a mean of \SI{-12}{\percent} with a range of \qtyrange{-41}{10}{\percent} for the left parotid and a mean of \SI{-3941}{\mm^3} with a range of \qtyrange{-29112}{4584}{\mm^3} corresponding to a mean of \SI{-12}{\percent} with a range of \qtyrange{-41}{10}{\percent} for the right parotid. This is slightly smaller but in line with previous studies, considering the averaging effect caused by the pCT-rCT pairings from the training set. The study of \citep{santosMorphologyVolumeDensity2020a} reports a mean of \SI{-6318}{\mm^3} with a range of \qtyrange{-23883}{6033}{\mm^3} or \SI{-19.8}{\percent} with a range of \qtyrange{-61}{21.1}{\percent} for the left parotid and a mean of \SI{-6809}{\mm^3} with a range of \qtyrange{-21744}{6121}{\mm^3} or \SI{-21}{\percent} with a range of \qtyrange{-59.3}{28.5}{\percent} for the right parotid. The study of \citep{vasquezosorioLocalAnatomicChanges2008} reports an average volume reduction of \SI{14}{\percent} and the study of \citep{bhideWeeklyVolumeDosimetric2010} reports \SI{14}{\percent} or \SI{16}{\percent}. 

The COM shifts observed in the dataset are a median of \SI{2}{\mm} with a range of \qtyrange{0.2}{13}{\mm} for the left parotid and a median of \SI{3}{\mm} with a range of \qtyrange{0.4}{12}{\mm} for the right parotid. These values are in agreement with the median of \SI{3.1}{\mm} in a range of \qtyrange{0}{9.9}{\mm} reported by \citep{barkerQuantificationVolumetricGeometric2004}.

To conclude, the distributions from the training set are deemed in line with the expectations set out by previous studies. Differences between the data presented here and the one from previous studies, such as \citep{medberyVariationParotidGland2000a} and \citep{santosMorphologyVolumeDensity2020a} can be attributed to several factors. First, the pCT-rCT composition of the training set is bound to underestimate the changes when compared to studies based on only pCT-final CT pairs. Second, differences are expected due to the anonymization of the training set and the differences between the compared cohorts. Previous studies such as the ones of \citep{ericsonNormalVariationParotid1970,santosMorphologyVolumeDensity2020a,vasquezosorioLocalAnatomicChanges2008} showed differences in parotid volumes depending on age, sex, weight, smoker status, planned doses, degree of parotid sparing and treatment modality. Third, a small effect could be expected due to inter-observer variability and systematic errors introduced by interpolating the original images on a new, coarser grid could also influence the observed absolute volumes.

\subsection{Generative performance}
\label{subsec:gen_performance}

To assess the generative performance of the model, the test set has been characterized in a similar way to the training set. Table \ref{tab:statistics_comparison} illustrates the statistics obtained from the test set. The values of the absolute and relative volumetric changes and of the COM shifts are in line with the ones obtained for the training set in Table \ref{tab:train_set_stats}. On the test set, the statistics are a relative volume difference of \SI{-12}{\percent} and an average COM shift of \SI{2}{\mm} for the left parotid, and an average relative difference of \SI{-11}{\percent} and an average COM shift of \SI{3}{\mm} for the right parotid.

The test set, that contained 9 patients, was input into the final trained model and 100 samples were drawn for each record (pair of pCT-rCT) in the test set. The statistics on a parotid level can also be seen in Table \ref{tab:statistics_comparison}. The values that the model predicts are close to the ones obtained from the test set, with an average relative difference of \SI{-15}{\percent} and an average COM shift of \SI{3}{\mm} for the left parotid, and an average relative difference of \SI{-11}{\percent} and an average COM shift of \SI{3}{\mm} for the right parotid. 

\begin{table}[H]
    \centering
    \resizebox{\textwidth}{!}{%
    \begin{tabular}{|cll|lllll|}
    \hline
    \multicolumn{3}{|c|}{}                                                                                             & \multicolumn{5}{c|}{Statistic}         \\ \hline
    \multicolumn{1}{|c|}{Organ} &
      \multicolumn{1}{c|}{Set} &
      \multicolumn{1}{c|}{Metric} &
      \multicolumn{1}{c}{Mean} &
      \multicolumn{1}{c}{SD} &
      \multicolumn{1}{c}{Min.} &
      \multicolumn{1}{c}{Median} &
      \multicolumn{1}{c|}{Max.} \\ \hline
    \multicolumn{1}{|c|}{\multirow{9}{*}{Parotid L}} & \multicolumn{1}{l|}{Test}      & Planning volume (\unit{\mm^3}) & 35773 & 12002 & 15704  & 33112 & 52312 \\
    \multicolumn{1}{|c|}{}                           & \multicolumn{1}{l|}{Test}      & Repeat volume (\unit{\mm^3})   & 31186 & 11012 & 13568  & 29312 & 49080 \\
    \multicolumn{1}{|c|}{}                           & \multicolumn{1}{l|}{Generated} & Repeat volume (\unit{\mm^3})   & 30870 & 9730  & 11968  & 30280 & 52216 \\
    \multicolumn{1}{|c|}{}                           & \multicolumn{1}{l|}{Test}      & Difference (\unit{\mm^3})      & -4587 & 3937  & -15336 & -3232 & 392   \\
    \multicolumn{1}{|c|}{}                           & \multicolumn{1}{l|}{Generated} & Difference (\unit{\mm^3})      & -5636 & 3808  & -17080 & -5268 & 3560  \\
    \multicolumn{1}{|c|}{}                           & \multicolumn{1}{l|}{Test}      & Relative difference (\%)       & -12   & 9     & -33    & -10   & 1     \\
    \multicolumn{1}{|c|}{}                           & \multicolumn{1}{l|}{Generated} & Relative difference (\%)       & -15   & 8     & -47    & -16   & 11    \\
    \multicolumn{1}{|c|}{}                           & \multicolumn{1}{l|}{Test}      & COM shift (\unit{\mm})         & 2     & 1     & 0.6    & 2     & 6     \\
    \multicolumn{1}{|c|}{}                           & \multicolumn{1}{l|}{Generated} & COM shift (\unit{\mm})         & 3     & 1     & 5e-02  & 3     & 6     \\ \hline
    \multicolumn{1}{|c|}{\multirow{9}{*}{Parotid R}} & \multicolumn{1}{l|}{Test}      & Planning volume (\unit{\mm^3}) & 31786 & 10827 & 14200  & 32224 & 50784 \\
    \multicolumn{1}{|c|}{}                           & \multicolumn{1}{l|}{Test}      & Repeat volume (\unit{\mm^3})   & 28196 & 10098 & 13168  & 25256 & 48288 \\
    \multicolumn{1}{|c|}{}                           & \multicolumn{1}{l|}{Generated} & Repeat volume (\unit{\mm^3})   & 29938 & 11392 & 10144  & 27352 & 54456 \\
    \multicolumn{1}{|c|}{}                           & \multicolumn{1}{l|}{Test}      & Difference (\unit{\mm^3})      & -3589 & 3276  & -11840 & -2232 & 784   \\
    \multicolumn{1}{|c|}{}                           & \multicolumn{1}{l|}{Generated} & Difference (\unit{\mm^3})      & -3195 & 2642  & -10240 & -2916 & 3672  \\
    \multicolumn{1}{|c|}{}                           & \multicolumn{1}{l|}{Test}      & Relative difference (\%)       & -11   & 10    & -32    & -7    & 6     \\
    \multicolumn{1}{|c|}{}                           & \multicolumn{1}{l|}{Generated} & Relative difference (\%)       & -11   & 8     & -32    & -10   & 9     \\
    \multicolumn{1}{|c|}{}                           & \multicolumn{1}{l|}{Test}      & COM shift (\unit{\mm})         & 3     & 2     & 0.5    & 2     & 6     \\
    \multicolumn{1}{|c|}{}                           & \multicolumn{1}{l|}{Generated} & COM shift (\unit{\mm})         & 3     & 1     & 3e-01  & 2     & 7     \\ \hline
    \end{tabular}%
    }
    \caption{Comparison of statistics based on the test and generated (with 100 samples per record) sets. The table displays for both parotid glands per set the mean, standard deviation, minimum, median and maximum of the volume on the planning and repeat CT images, the difference between these volumes (absolute and relative) and the center of mass shifts. }
    \label{tab:statistics_comparison}
\end{table}

\begin{figure}[H]
    \centering
    \begin{subfigure}{\linewidth}
        \includegraphics[width=\columnwidth]{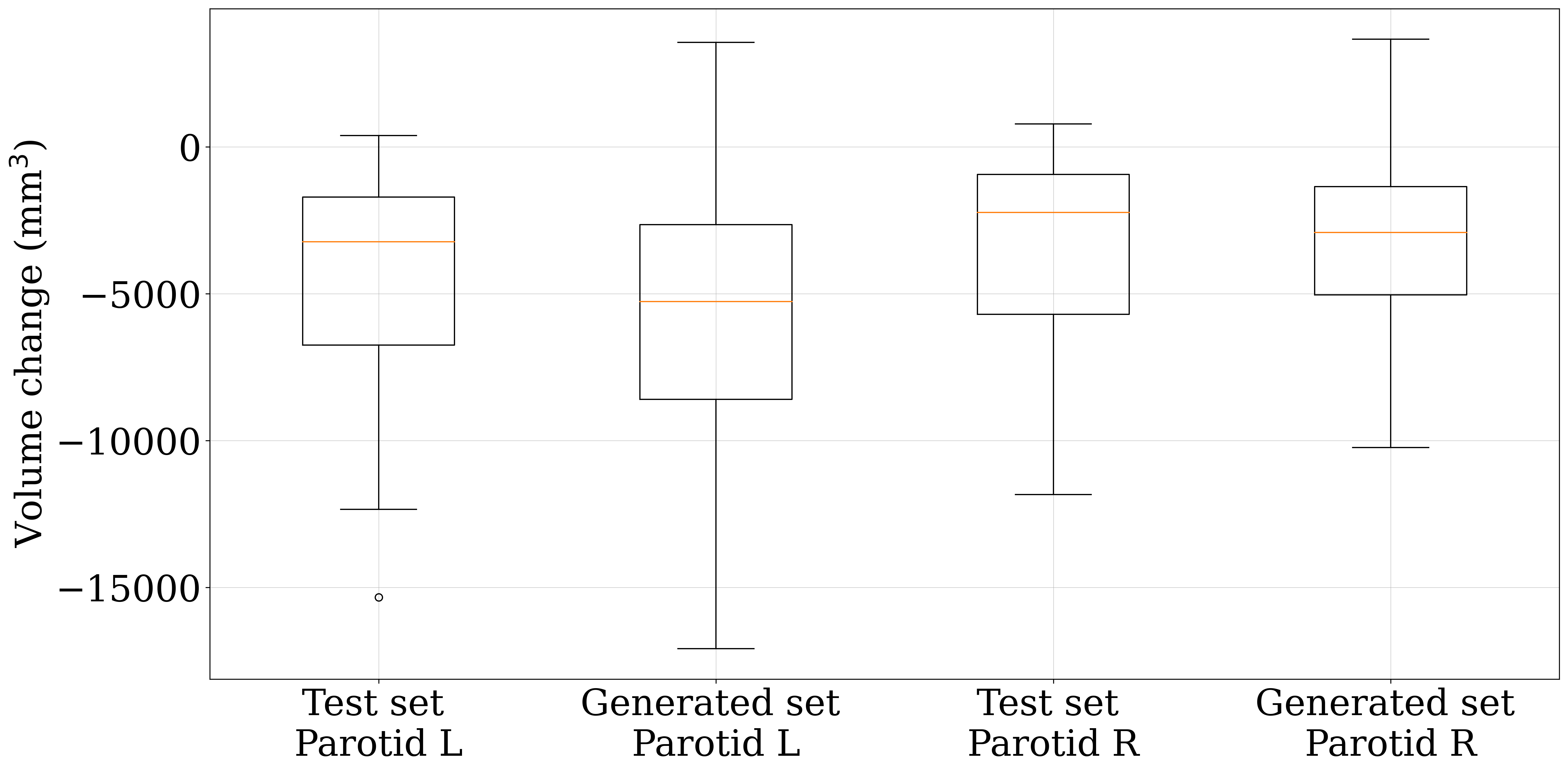}
        \caption{Organ specific box plot of absolute difference in organ volume changes.}
        \label{fig:oar_boxplot_volume_change}
    \end{subfigure}
    \begin{subfigure}{\linewidth}
        \includegraphics[width=\columnwidth]{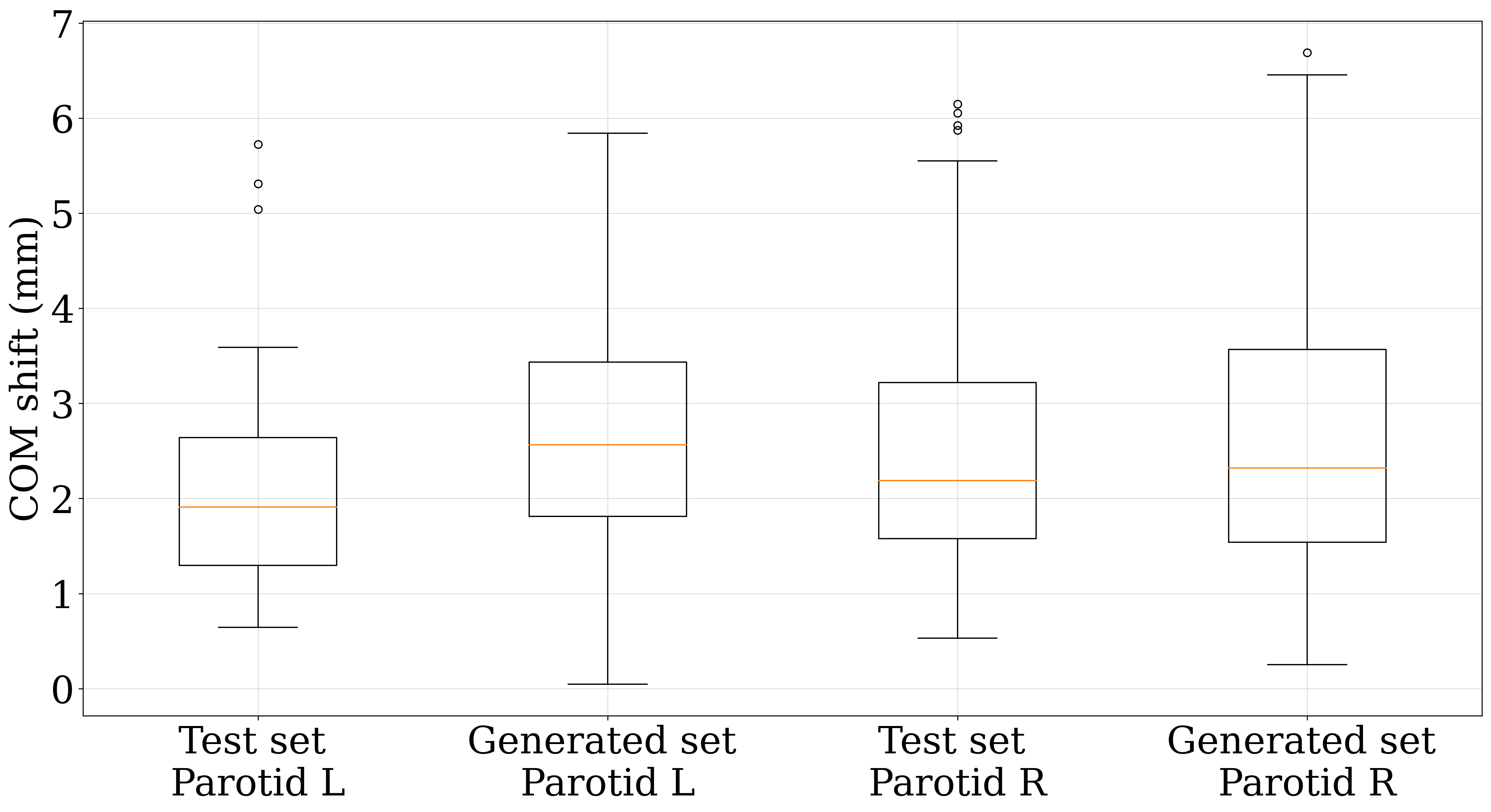}
        \caption{Organ specific box plot of COM shifts.}
        \label{fig:oar_boxplot_com_shifts}
    \end{subfigure}
    \caption{Organ specific generative performance. The figures display boxplots with the $x$ axis showing the organ and the $y$ axis giving either the absolute volumetric changes or the COM shifts for the given organ.}
    \label{fig:gen_set_oar_boxplot}
\end{figure}

The comparison of the statistics from Tables \ref{tab:statistics_comparison} is visualized in Figure \ref{fig:gen_set_oar_boxplot}. Figure \ref{fig:oar_boxplot_volume_change} displays for both parotids boxplots of the volume changes on both the test and generated set, while Figure \ref{fig:oar_boxplot_com_shifts} displays for both parotids boxplots of the COM shifts on both the test and generated set. Both figures show that the model is capable of capturing the anatomical changes observed in the test set. With the exception of the minimum volume generated for the right parotid, the model is capable of generating distributions broad enough to capture the real ones. The medians of the distributions are also in close agreement, with some discrepancy observed for the left parotid. Thus, given Table \ref{tab:statistics_comparison} and Figure \ref{fig:gen_set_oar_boxplot} it can be concluded that the model is capable of accurately modelling the anatomical changes present in the test set. 

\begin{figure}[H]
    \centering
    \includegraphics[width=\linewidth]{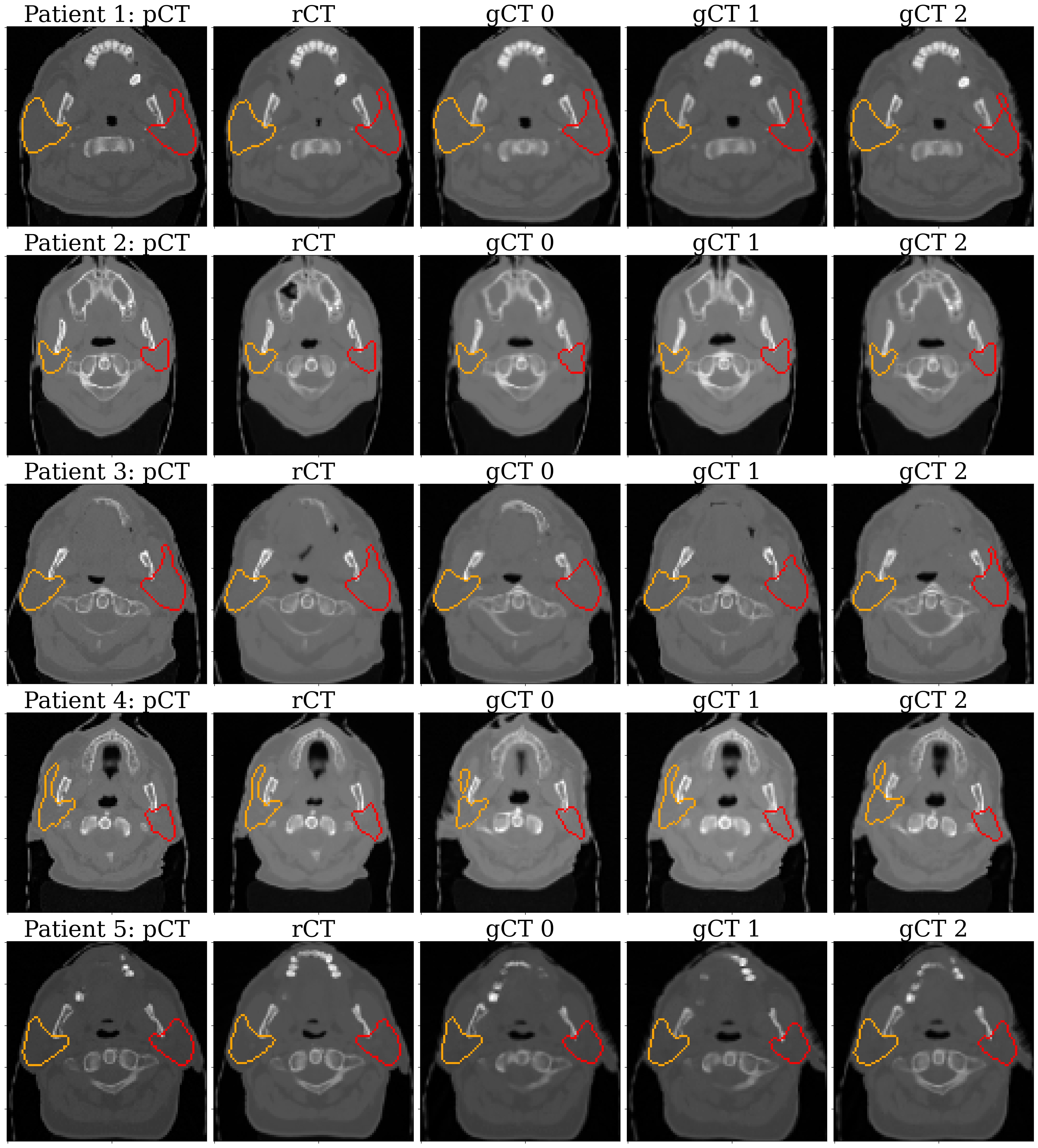}
    \caption{Example of generated images. The figure displays, for 5 randomly selected patients from the test set, in the first column the true pCT, in the second column one of the true rCTs and in the remaining columns generated CT images.}
    \label{fig:gen_imgs}
\end{figure}

An illustration of the generative capabilities of the model is shown in Figure \ref{fig:gen_imgs}. The figure displays for 5 patients in the test set, in the first column the pCT, in the second column one of the rCTs and in the following 3 columns three patient specific generated CT images with corresponding contours (the left parotid colored in red and the right parotid colored in orange). As already mentioned in Table \ref{tab:literature_anatomical_changes}, the flattening and medial movement of the parotids is expected. This feature is clearly illustrated for the third patient in the generated images shown in columns \numrange{3}{5}. Moreover, the model also predicts neck pose shifts, as illustrated by the smaller air gap in the oral cavity of the first generated CT of patient 4 or by the change in the shown dentition of patient 5. Overall, the figure supports the conclusion that the model is capable of generating realistic anatomies that involve posture shifts, shifting air gaps and the typical expected anatomical changes in the parotid glands. 

\begin{figure}[H]
    \centering
    \begin{subfigure}{.9\linewidth}
        \includegraphics[width=\columnwidth]{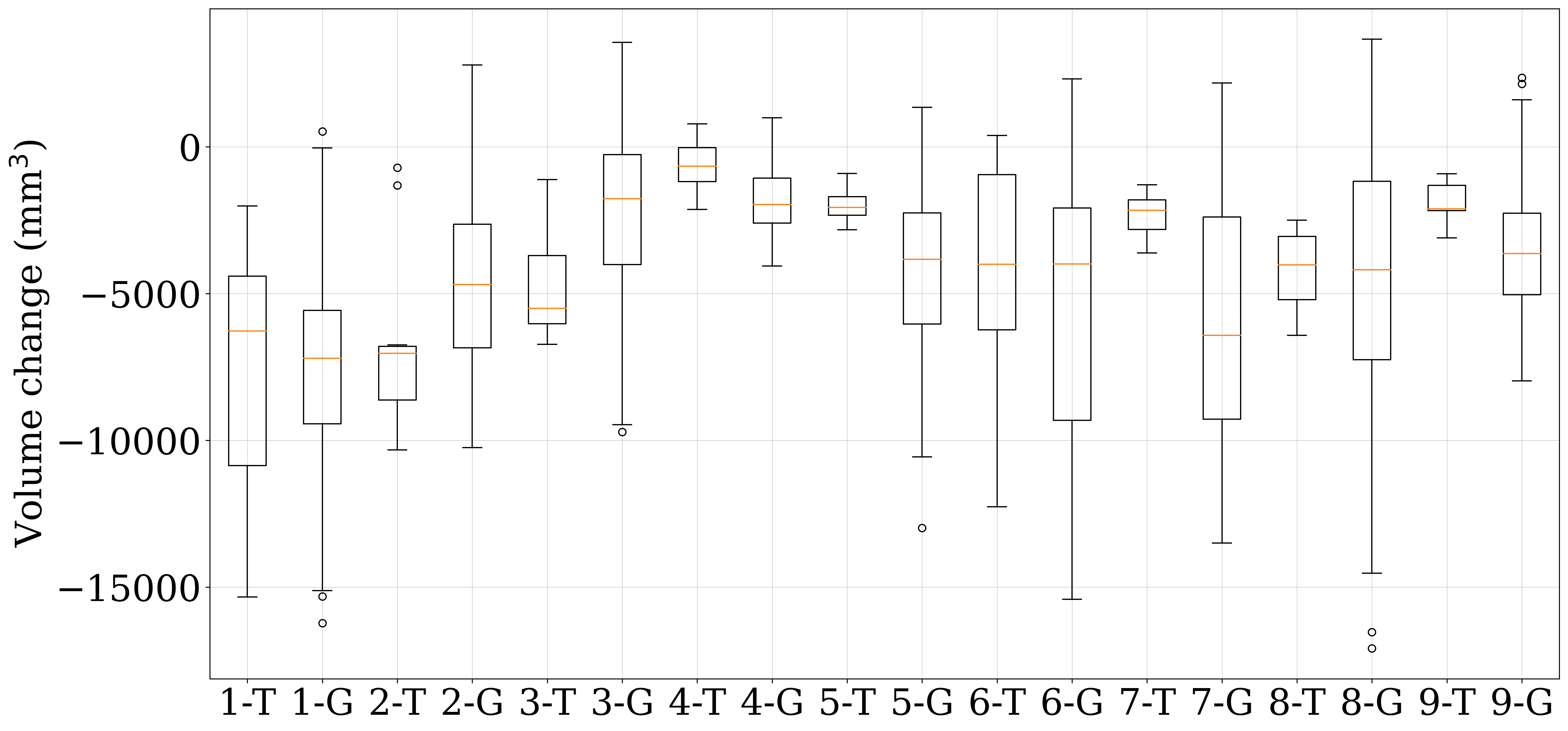}
        \caption{Patient specific box plot of absolute difference in organ volume changes.}
        \label{fig:pat_boxplot_volume_change}
    \end{subfigure}
    \begin{subfigure}{.9\linewidth}
        \includegraphics[width=\columnwidth]{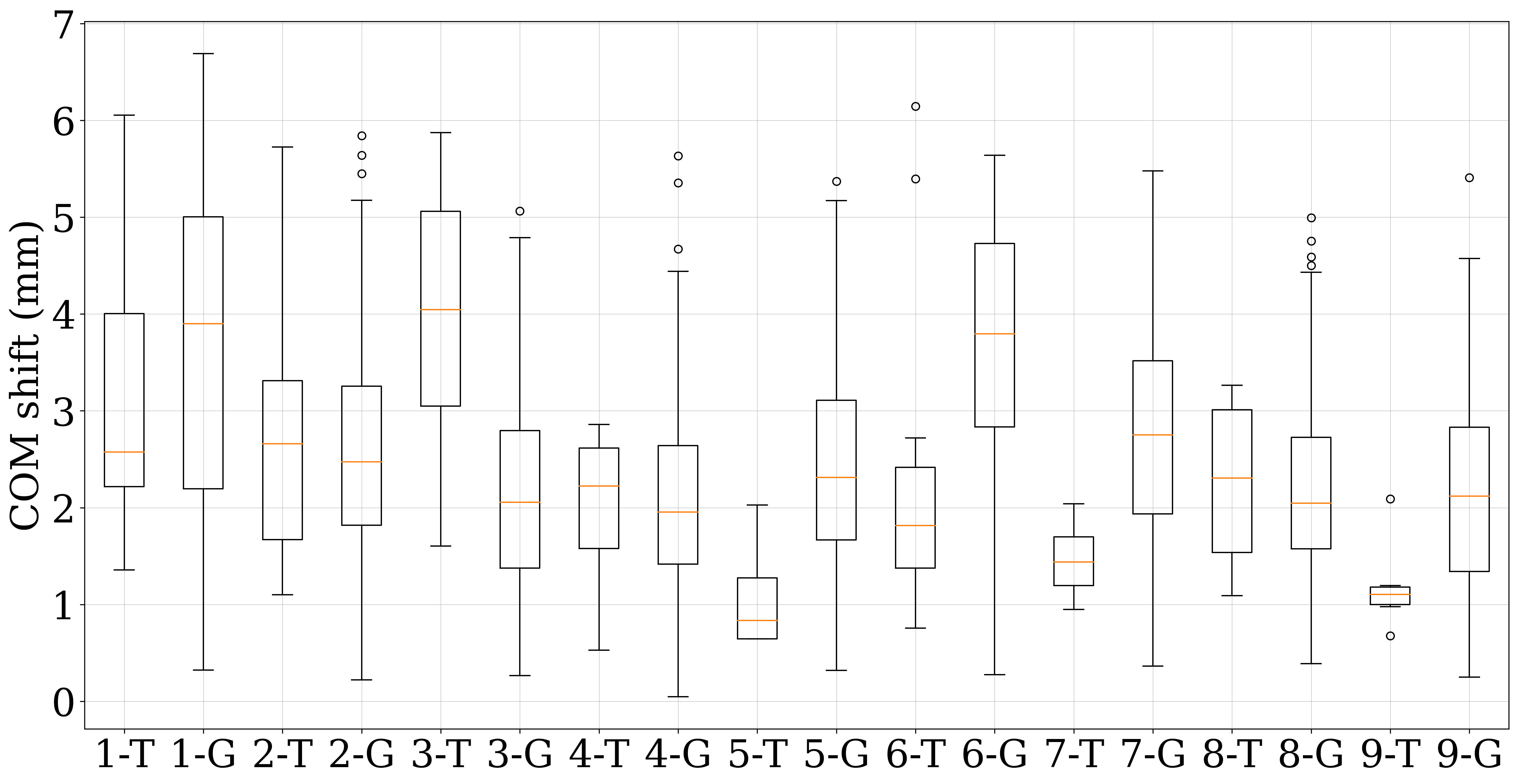}
        \caption{Patient specific box plot of COM shifts.}
        \label{fig:pat_boxplot_com_shifts}
    \end{subfigure}
    \caption{Patient specific generative performance. The figures display boxplots with the $x$ axis showing the organ and the $y$ axis giving either the absolute volumetric changes or the COM shifts for the given organ.}
    \label{fig:gen_set_pat_boxplot}
\end{figure}

To further test the population based model, Figure \ref{fig:gen_set_pat_boxplot} shows patient-specific boxplots of the anatomical changes. Figure \ref{fig:pat_boxplot_volume_change} displays for each patient in the test set, the true volumetric change (denoted by the patient number and -T) and the generated volumetric changes by drawing 100 samples (denoted by the patient number and -G). In the case of both the volumetric changes illustrated in Figure \ref{fig:pat_boxplot_volume_change} and the COM shifts illustrated in Figure \ref{fig:pat_boxplot_com_shifts}, the model largely predicts broad enough distributions that encompass the true ones. The agreement of the true and generated medians, however, varies per patient. The model is capable of producing accurate patient-specific anatomical changes for patients 4 and 8, and less so for patients 3 and 7. This likely illustrates the non-patient specific nature of the model coupled with an insufficient number of recorded repeat CT images for those patients. Similar conclusions can be drawn for the patient-specific COM shifts displayed in Figure \ref{fig:pat_boxplot_com_shifts}

\subsection{Comparison to DiffuseRT}
\label{subsec:diffuse_rt}

\begin{figure}
    \centering
    \begin{subfigure}{.9\linewidth}
        \includegraphics[width=\columnwidth]{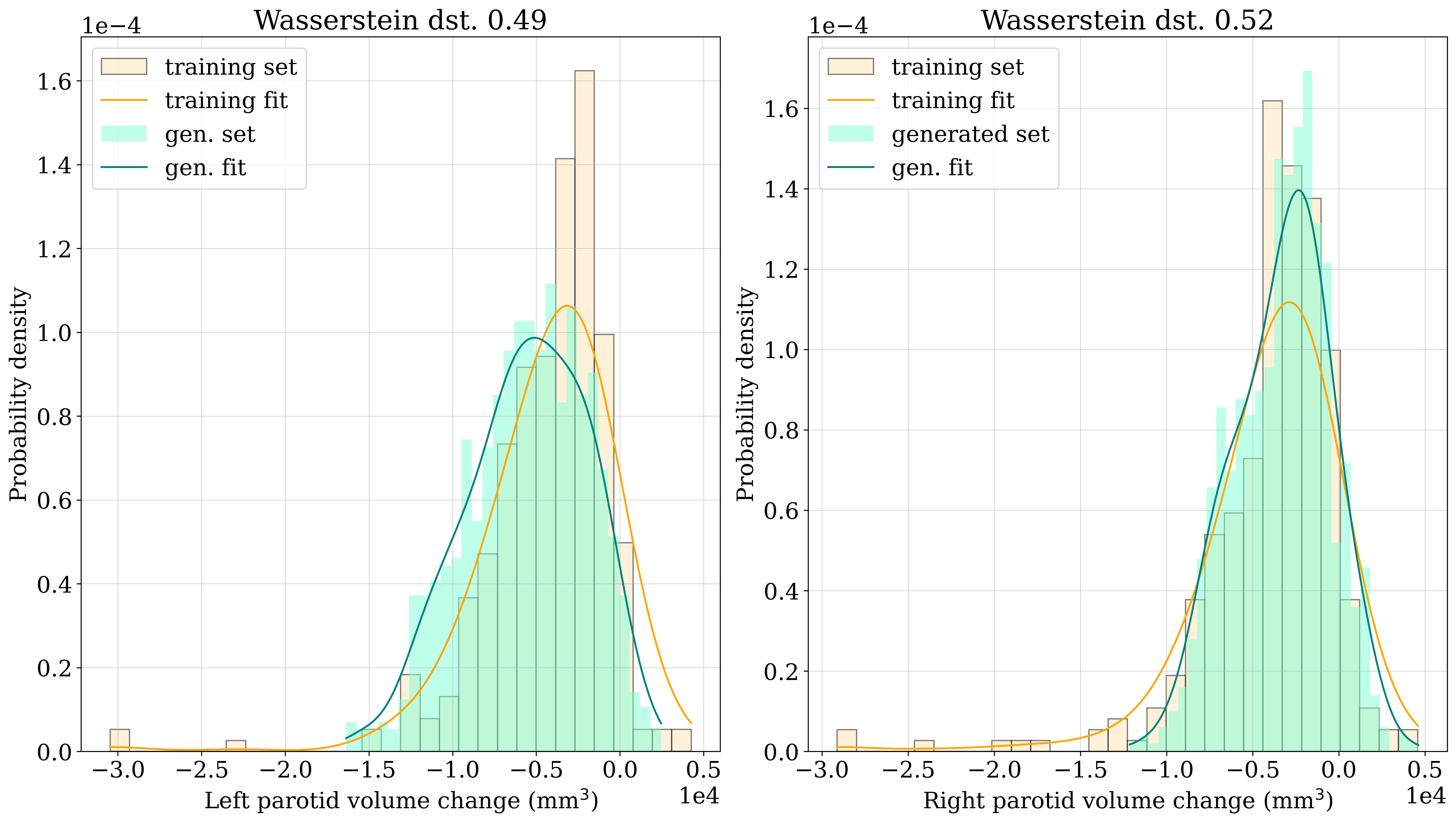}
        \caption{Left and right parotid volume change distributions and corresponding kernel density estimate for both the true and generated sets. The Wasserstein distance between the distributions is given in the title.}
        \label{fig:wasserstein_dists_rel_delta}
    \end{subfigure}
    \begin{subfigure}{.9\linewidth}
        \includegraphics[width=\columnwidth]{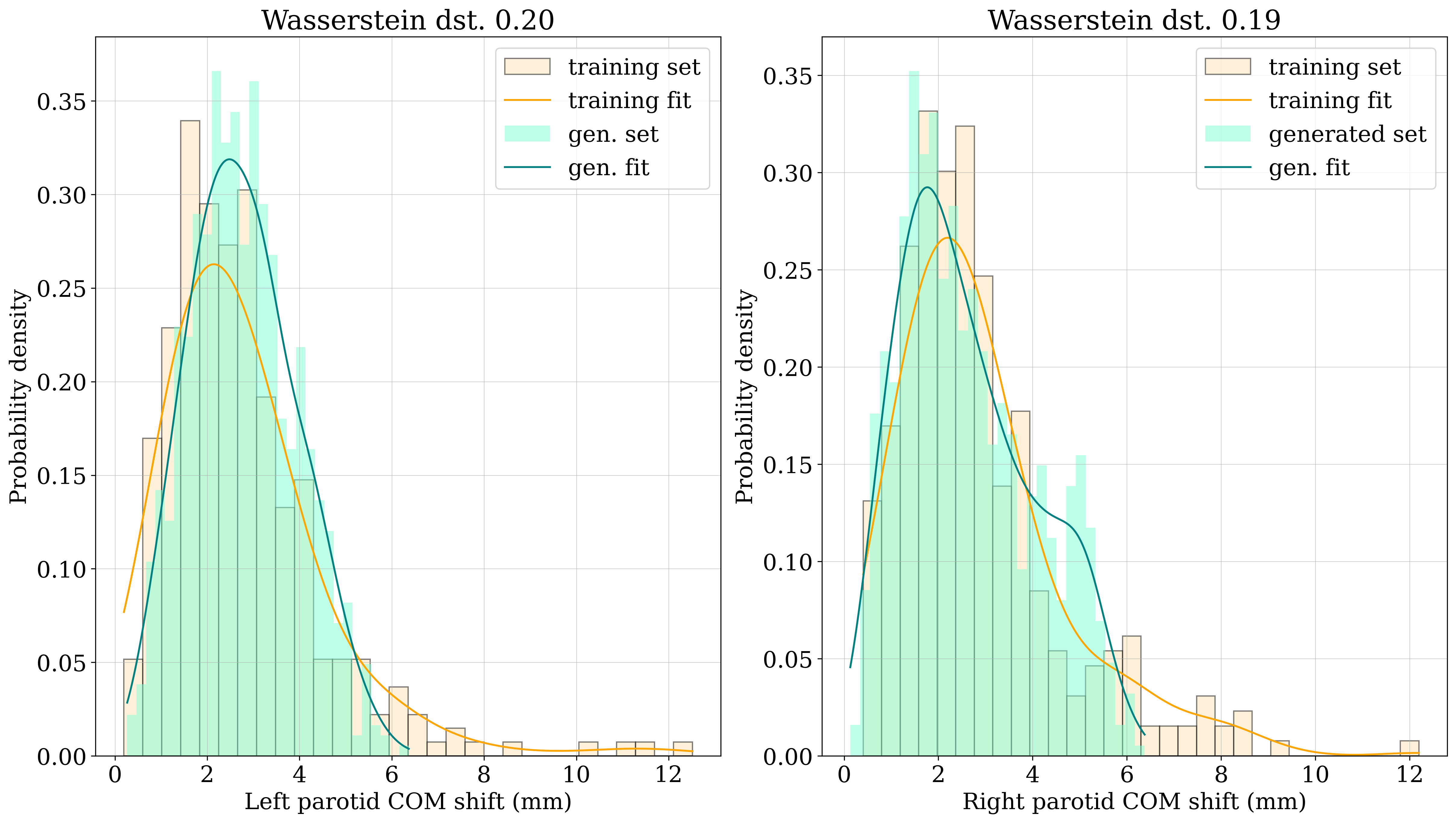}
        \caption{Left and right parotid COM shift distributions and corresponding kernel density estimate for both the true and generated sets. The Wasserstein distance between the distributions is given in the title.}
        \label{fig:wasserstein_dists_COM_shift}
    \end{subfigure}
    \caption{Comparison of true and generated anatomical change distributions. The figures display for both left and right parotids the true and generated anatomical change distributions and their corresponding kernel density estimates.}
    \label{fig:wasserstein_dists}
\end{figure}

The generative performance of DAM with respect to principal component analysis (PCA) based models has already been documented in the previous work of \citep{pastor-serranoProbabilisticDeepLearning2023}, where it was shown to outperform them. Thus, the generative performance of this model was compared with the recently published denoising diffusion probabilistic model (DDPM) of \citep{smoldersDiffuseRTPredictingLikely2024}. DDPM is also a generative deep learning model that approximates a data distribution, by inverting a gradual multi-step noise addition process. Similarly to the results shown by DDPM, Figure \ref{fig:wasserstein_dists} displays for each parotid, the true (training set) and generated volume change distributions (in Figure \ref{fig:wasserstein_dists_rel_delta}) and COM shift distribution (in Figure \ref{fig:wasserstein_dists_COM_shift}) together with a kernel density estimate for each. The kernel density estimate was computed using the Scikit library \citep{pedregosaScikitlearnMachineLearning2011} and a kernel bandwidth defined as one tenth of the range of values in the distribution. Both volume change and COM shift distributions that the DAM\textsubscript{HN} training set exhibits are qualitatively different than the ones reported by DDPM, displaying less bimodality. This difference is likely attributable to the difference in set sizes (9 patients for DDPM versus 83 patients for DAM\textsubscript{HN}). The kernel density estimates for the training and generated sets are generally in agreement, with disagreement occurring at the ends of the distributions, as is visible in Figure \ref{fig:wasserstein_dists_COM_shift}.  

DAM\textsubscript{HN} and DDPM were also compared in terms of the Wasserstein distance (WD) between the true (training set) and generated anatomical changes distributions. The Wasserstein distance is a metric for probability distribution similarity, with a value of zero occurring when the distributions are the same and larger values indicating more different distributions. To compute it, the volume changes in the parotids and the COM shifts were input into the SciPy implementation \citep{virtanenSciPyFundamentalAlgorithms2020} together with the corresponding weights, after conversion into histogram values. As the Wasserstein distance exhibits a scaling effect based on the range of the data, the Wasserstein distances were normalized by the average of the changes on the training and generated sets. Table \ref{tab:wasserstein_comparison} shows the comparison between DDPM and DAM\textsubscript{HN}. The qualitative agreement observed in Figure \ref{fig:wasserstein_dists} is also illustrated by the low Wasserstein distance, which is comparable between DDPM and DAM\textsubscript{HN} for all metrics. 

\begin{table}[H]
    \centering
    \resizebox{0.7\linewidth}{!}{%
    \begin{tabular}{|l|l|cc|}
    \hline
    \multicolumn{1}{|c|}{\multirow{2}{*}{Metric}} & \multicolumn{1}{c|}{\multirow{2}{*}{Structure}} & \multicolumn{2}{c|}{Model} \\ \cline{3-4} 
    \multicolumn{1}{|c|}{} & \multicolumn{1}{c|}{} & \multicolumn{1}{c|}{DDPM} & DAM\textsubscript{HN} \\ \hline
    \multirow{2}{*}{$\Delta$ Volume} & Left parotid & \multicolumn{1}{c|}{0.25} & 0.49 \\
     & Right parotid & \multicolumn{1}{c|}{0.41} & 0.52 \\ \hline
    \multirow{2}{*}{COM shift} & Left parotid & \multicolumn{1}{c|}{0.22} & 0.20 \\
     & Right parotid & \multicolumn{1}{c|}{0.19} & 0.19 \\ \hline
    \end{tabular}%
    }
    \caption{Wasserstein distance comparison between the DDPM model of \citep{smoldersDiffuseRTPredictingLikely2024} and DAM\textsubscript{HN}. The table displays the Wasserstein distance between the true (training set) and generated volume loss and COM shift distributions in the left and right parotids.}
    \label{tab:wasserstein_comparison}
\end{table}

\subsection{Latent space analysis}
\label{subsec:latent_space_analysis}

\begin{figure}[H]
    \centering
    \begin{subfigure}{\linewidth}
        \includegraphics[width=\columnwidth]{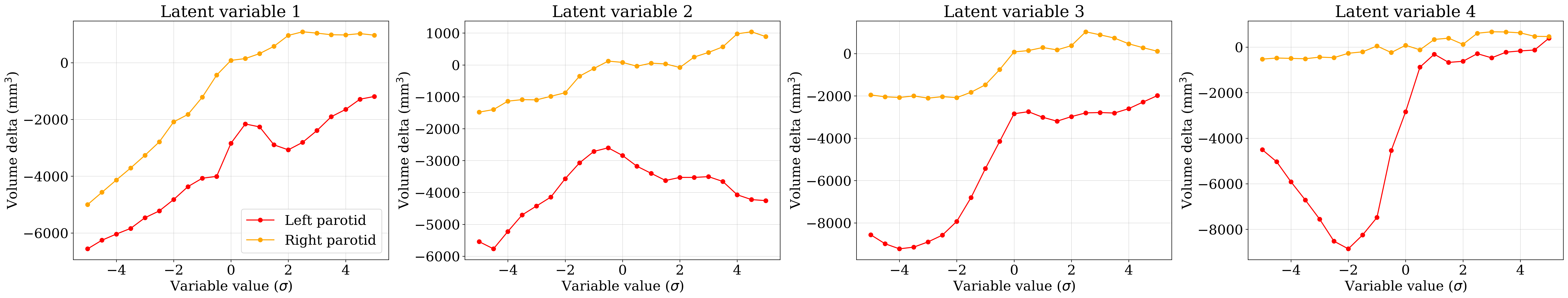}
        \caption{Parotid volume changes for each latent variable}
        \label{fig:latent_space_vol_changes}
    \end{subfigure}
    \begin{subfigure}{\linewidth}
        \includegraphics[width=\columnwidth]{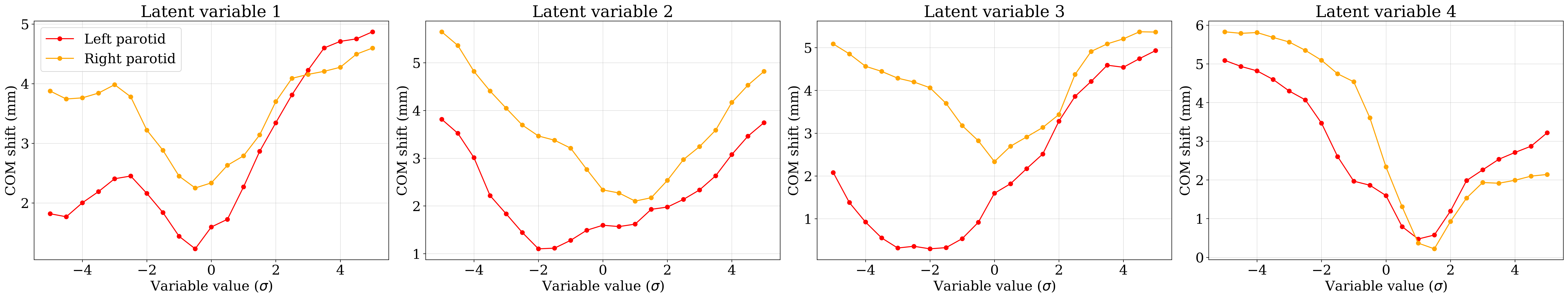}
        \caption{Parotid COM shifts for each latent variable}
        \label{fig:latent_space_COM_changes}
    \end{subfigure}
    \caption{Latent space variations. For a model with 4 latent variables, the two figures display the parotid volume (first row) and COM changes (second row) that individual latent variables cause. The latent variable was varied from $-5 \sigma$ to $5 \sigma$ while the remaining variables were set to 0.}
    \label{fig:latent_space_analysis}
\end{figure}

To facilitate ease of plotting, this subsection details a latent space analysis for a model with 4 latent variables and a randomly chosen patient in the test set. Figure \ref{fig:latent_space_analysis} illustrates the anatomical variations that occur when an individual latent variable is varied from  $-5 \sigma$ to $5 \sigma$, while the others are kept fixed to 0. 

\begin{figure}[H]
    \centering
    \includegraphics[width=\linewidth]{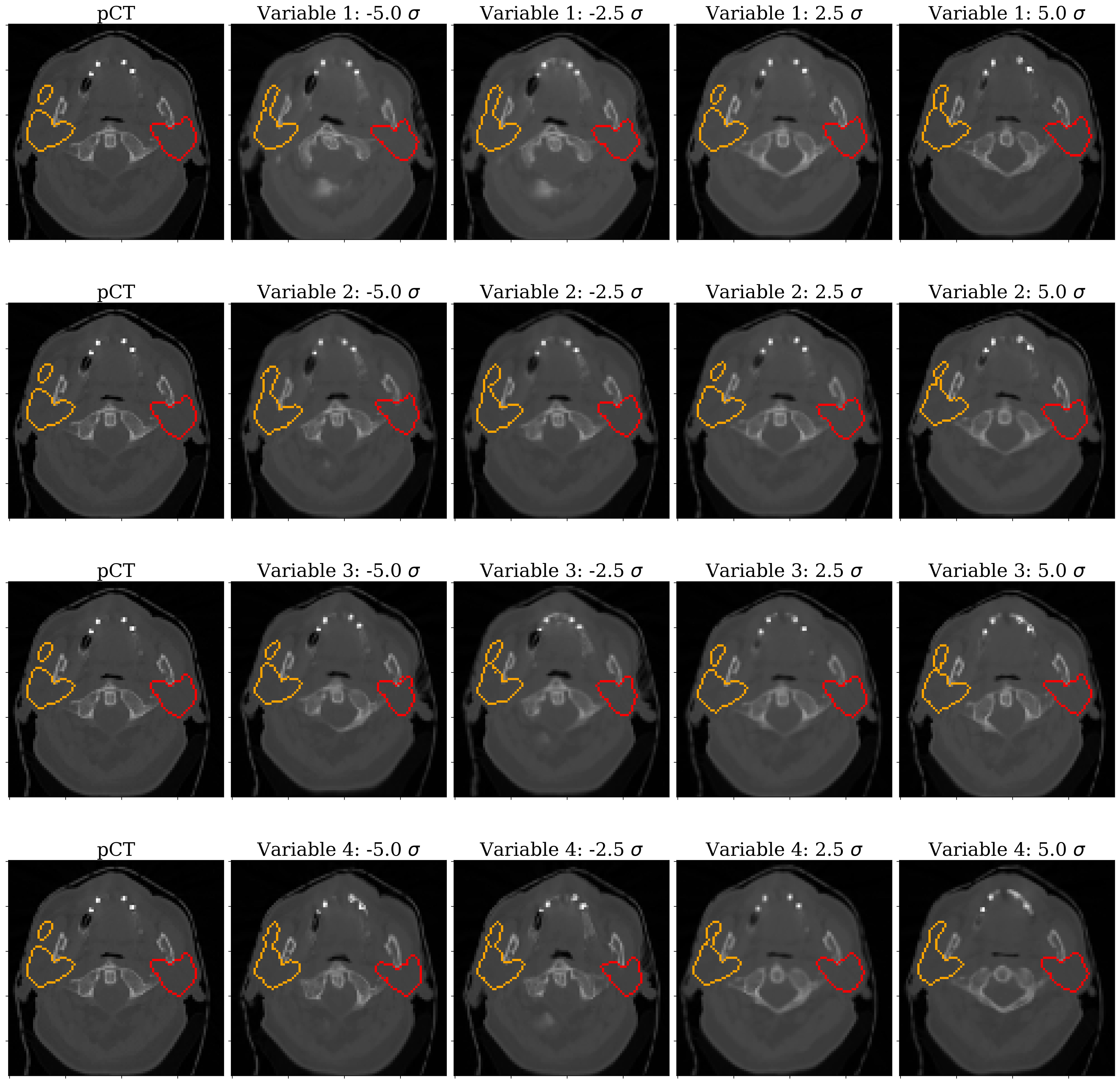}
    \caption{Latent space visualization. The figures display in the first column for a given patient, the pCT and associated parotids. In the following columns the figure displays the latent variable number and the value it was set to.}
    \label{fig:latent_space_imgs}
\end{figure}

Figure \ref{fig:latent_space_vol_changes} shows consistently larger volumetric lossess in the left parotid in comparison to the right parotid. Given that previous studies recorded larger lossess in irradiated parotids as opposed to non-irradiated ones \citep{vasquezosorioLocalAnatomicChanges2008}, this result could indicate the increased dose delivered to the left parotid for this patient cohort. Also visible in Figure \ref{fig:latent_space_vol_changes} is the relatively smooth latent space that the model learns. Figure \ref{fig:latent_space_COM_changes} shows that for both parotids, the COM deformations are roughly similar in absolute value. This is in line with the expectation, set by the work of \citep{vasquezosorioLocalAnatomicChanges2008}, that both parotids move in the medial direction with similar amplitudes. Moreover, Figure \ref{fig:latent_space_COM_changes} also shows that similar latent variables produce similar behaviors in both parotids and that the learned latent space is smooth. Volume and COM shifts are just one measure of latent space variations. Figure \ref{fig:latent_space_imgs} shows, for a patient in the test set, the images produced when individual latent variables are varied. The first column of the figure displays the pCT, while the remaining columns display the image and associated contours created by the individual latent variables (with its value given in the title of the figure). In this case, it appears that variable 1 results in neck pose shifts, identified by the changing esophagus air gaps as the latent variable is varied. 

\section{Conclusion}
\label{sec:conclusion}

This work presented a probabilistic deep learning model for generating future anatomical changes in H\&N radiotherapy patients. The model was trained on a training set coming from 83 patients and was assessed on test set coming from 9 patients. On the test set the model achieved a DICE score of \num{0.92} and an NCC score of \num{0.63} using 32 latent variables. The model produces volumetric changes and COM shift distributions that are broad enough to capture the real, observed ones, with the predicted means being close to the real ones. On the test set the mean left parotid volumetric loss was \SI{12}{\percent} while the model predicted \SI{15}{\percent}, and the mean right parotid volumetric loss was \SI{11}{\percent} while the model predicted \SI{11}{\percent}. On the test set, the mean COM shift in the left parotid was \SI{2}{\mm} while the model predicted \SI{3}{\mm} and for the right parotid it was \SI{3}{\mm} while the model predicted \SI{3}{\mm}. DAM\textsubscript{HN} was compared to the state of the art denoising diffusion probabilistic model (DDPM) for H\&N anatomical changes presented by \citep{smoldersDiffuseRTPredictingLikely2024}. For both parotid glands, DAM\textsubscript{HN} achieved similar Wasserstein distances to the ones obtained by the DDPM model between the true and generated volume loss distributions (\num{0.49} versus \num{0.25} and \num{0.52} versus \num{0.41})  and between the COM shift distributions (\num{0.20} versus \num{0.22} and \num{0.19} versus \num{0.19}). The latent space analysis showed that the model learns a smooth latent space, that displays some correlation between the latent variables (which was not discouraged in the model framework). The model, and the comparison of DAM\textsubscript{HN} to the model of \citep{smoldersDiffuseRTPredictingLikely2024}, could be improved by adding more of the available RT structures to the datasets. The same ones that DDPM used (body and esophagus) and additional ones (e.g., target, submandibular glands, oral cavity) could improve the quality of the generated anatomies. Moreover, the inclusion of these additional RT structures would create additional metrics to test against the literature reported ones and provide a more complete generative performance assessment. Additionally, architectural details such as the number of layers and the optimal hyperparameters should be investigated in more detail. Overall, DAM\textsubscript{HN} was capable of quickly generating hundreds of realistic images of inter-fractional anatomies. As already mentioned, such a model has a number of applications in the radiotherapy workflow, such as improving robust optimization, as a component in plan quality assurance in online adaptive proton therapy or in expanding the plan library approach. 

\section{Conflicts of interest and acknowledgements}

The authors wish to acknowledge that the manuscript is partly funded by Varian, a Siemens Healthineers Company.
The authors declare that they have no known competing financial interests or personal relationships that could have appeared to influence the work reported in this paper. Moreover, no data was used for the research described in the article.

\section{Credit statement}

\textbf{Tiberiu Burlacu:} Conceptualization, methodology, software, validation, formal analysis, data curation, investigation, writing - original draft, writing - review \& editing, visualization. \\
\textbf{Zolt\'{a}n Perk\'{o}:} Conceptualization, methodology, validation, resources, writing - review \& editing, supervision, project administration, funding acquisition. \\
\textbf{Danny Lathouwers:} resources, writing - review \& editing, supervision. \\
\textbf{Mischa Hoogeman:} resources, writing - review \& editing.

\appendix

\bibliographystyle{agsm}
\bibliography{./bib/bibliography.bib}

\end{document}